\documentclass[prd,aps,twocolumn,a4paper,showkeys,nofootinbib]{revtex4-1}

\usepackage{graphicx,psfrag}
\usepackage{mathrsfs}
\usepackage{amsmath,amsfonts,amssymb}
\usepackage{multirow}
\usepackage{comment}
\usepackage[T1]{fontenc}

\usepackage{ulem}
\usepackage{hyperref}
\usepackage{enumitem}
\usepackage{accents}

\newcommand{\be}{\begin{equation}}
\newcommand{\ee}{\end{equation}}
\newcommand{\bea}{\begin{eqnarray}}
\newcommand{\eea}{\end{eqnarray}}
\newcommand{\bel}{\begin{align}}
\newcommand{\eel}{\end{align}}

\def\lm{{\ell m}}

\def\GMc2{{\rm G M_{\odot} c^{-2}}}

\def\kt2{\kappa^\text{T}_2}

\usepackage{pifont} 

\newcommand{\teob}{\texttt{TEOBResumS}}

\newlength{\dhatheight}
\newcommand{\doublehat}[1]{%
    \settoheight{\dhatheight}{\ensuremath{\hat{#1}}}%
    \addtolength{\dhatheight}{-0.35ex}%
    \hat{\vphantom{\rule{1pt}{\dhatheight}}%
    \smash{\hat{#1}}}}

\usepackage{color}
\definecolor{cyan}{rgb}{0,0.9,0.9}
\definecolor{orange}{rgb}{0.9,0.5,0}
\definecolor{magenta}{rgb}{1,0,1}
\definecolor{purple}{rgb}{0.8,0.4,0.8}
\definecolor{gray}{rgb}{0.8242,0.8242,0.8242}

\begin{document}

\title{Numerical-relativity validation of effective-one-body waveforms \\in the intermediate-mass-ratio regime}

\author{Alessandro \surname{Nagar}${}^{1,2}$}
\author{James \surname{Healy}${}^3$}
\author{Carlos O. \surname{Lousto}${}^3$}
\author{Sebastiano \surname{Bernuzzi}${}^4$}
\author{Angelica \surname{Albertini}${}^{5,6}$}
\affiliation{${}^1$INFN Sezione di Torino, Via P. Giuria 1, 10125 Torino, Italy}
\affiliation{${}^2$Institut des Hautes Etudes Scientifiques, 91440 Bures-sur-Yvette, France}
\affiliation{${}^3$Center for Computational Relativity and Gravitation,
School of Mathematical Sciences,
Rochester Institute of Technology, 85 Lomb Memorial Drive, Rochester,
New York 14623, USA}
\affiliation{Theoretisch-Physikalisches Institut, Friedrich-Schiller-Universit{\"a}t Jena, 07743, Jena, Germany}
\affiliation{${}^5$Astronomical Institute of the Czech Academy of Sciences,
Bo\v{c}n\'{i} II 1401/1a, CZ-141 00 Prague, Czech Republic}
\affiliation{${}^6$Faculty of Mathematics and Physics, Charles University in Prague, 18000 Prague, Czech Republic}

\date{\today}

\begin{abstract}
One of the open problems in developing binary black hole (BBH) waveforms for gravitational wave astronomy is to model the
intermediate mass ratio regime and connect it to the extreme mass ratio regime. A natural approach is to employ the effective one body 
(EOB) approach to the two-body dynamics that, by design, can cover the entire mass ratio range and naturally incorporates the extreme mass ratio limit.
Here we use recently obtained numerical relativity (NR) data with mass ratios $m_1/m_2=(7,15,\,32,\,64,\,128)$ to test the accuracy 
of the state-of-the-art EOB model \teob{}  in the intermediate mass ratio regime. We generally find an excellent EOB/NR consistency 
around merger and ringdown for all mass ratios and for all available subdominant multipoles, except for the $\ell=m=5$ one.
This mode can be crucially improved using the new large-mass ratio NR data of this paper. The EOB/NR inspirals are also 
consistent with the estimated NR uncertainties.
We also use several NR datasets taken by different public catalogs to probe the universal behavior of the multipolar hierarchy 
of waveform amplitudes at merger, that smoothly connects the equal-mass BBH to the test-mass result.  
Interestingly, the universal behavior is strengthened if the nonoscillatory memory contribution is included in the NR waveform.
Future NR simulations with improved accuracy will be necessary to further probe, and possibly quantitatively refine, 
the \teob{} transition from late inspiral to plunge in the intermediate mass ratio regime.
\end{abstract}

\pacs{
  04.25.D-,     
  04.30.Db,   
  95.30.Sf,     
  %
  97.60.Lf    
}

\maketitle

\section{Introduction}

While ground based gravitational wave detectors like LIGO-Virgo  
\cite{LIGOScientific:2019fpa,Virgo} are particularly
sensitive to comparable (stellar) mass binaries,
third generation (3G) ground detectors~\cite{Purrer:2019jcp} and
space detectors, like LISA, will also be sensitive to the observation of very 
unequal mass binary black holes~\cite{Gair:2017ynp}. 
These will allow the search and study of intermediate mass black holes,
either as the large hole in a merger with a stellar mass black hole
(a source for 3G detectors) or
as the smaller hole in a merger with a supermassive black hole
(a source for LISA).

LISA will be sensitive to the inspiral and merger of black-hole
systems where the primary is significantly (10-1000 times) larger than
the secondary. These intermediate mass ratio inspiral (IMRI)
systems are crucial LISA sources, but their gravitational waveforms
are poorly understood, with component mass scales distinct enough to
present challenges for waveform modeling, particularly in numerical
simulations, and yet hard to match to black hole
perturbation theory computations in the extreme-mass-ratio inspirals
(EMRIs) regime.

The evolution of these large mass ratio
binaries has been approached via perturbation theory and the computation
of the gravitational self-force exerted by the field of the small 
black hole on itself~\cite{Pound:2015tma,Barack:2018yvs,Miller:2020bft,Hughes:2021exa,Pound:2021qin,Wardell:2021fyy,Warburton:2021kwk}.
The resolution of the binary black hole problem in its full nonlinearity
has been only possible after the 2005 breakthroughs in numerical 
relativity (NR)~\cite{Pretorius:2005gq,Campanelli:2005dd,Baker:2005vv},
and a first proof of principle has been performed in~\cite{Lousto:2010ut} 
for the 100:1 mass ratio case, following studies of 
the 10:1 and 15:1~\cite{Lousto:2010qx} ones.
In the case of~\cite{Lousto:2010ut} 
the evolution covered two orbits 
before merger, and while this proved that evolutions are possible,
practical application of these gravitational waveforms
requires longer evolutions. Other approaches to the large mass ratio
regime recently followed~\cite{Rifat:2019ltp,vandeMeent:2020xgc}.
A new set of evolutions that are based on the numerical techniques
refined for the longterm evolution of a spinning precessing binary with 
mass ratio $q\equiv m_1/m_2=15$~\cite{Lousto:2018dgd} have been used
in~\cite{Lousto:2020tnb} to perform a sequence of binary black hole simulations with increasingly
large mass ratios, reaching to a 128:1 binary that displays 13 orbits before
merger. Based on a detailed convergence study of the $q=15$ nonspinning
case, Ref.~\cite{Lousto:2020tnb} applied additional mesh refinements levels 
around the smaller hole horizon to reach successively the $q=32$, $q=64$, 
and $q=128$ cases. Reference~\cite{Lousto:2020tnb} also computed the
remnant properties of the merger as well as gravitational waveforms, peak frequency,
amplitude and luminosity. The obtained values were consistent with corresponding
phenomenological formulas, reproducing the particle limit within 2\%. 

\begin{figure*}[t]
	\center
\includegraphics[width=0.31\textwidth]{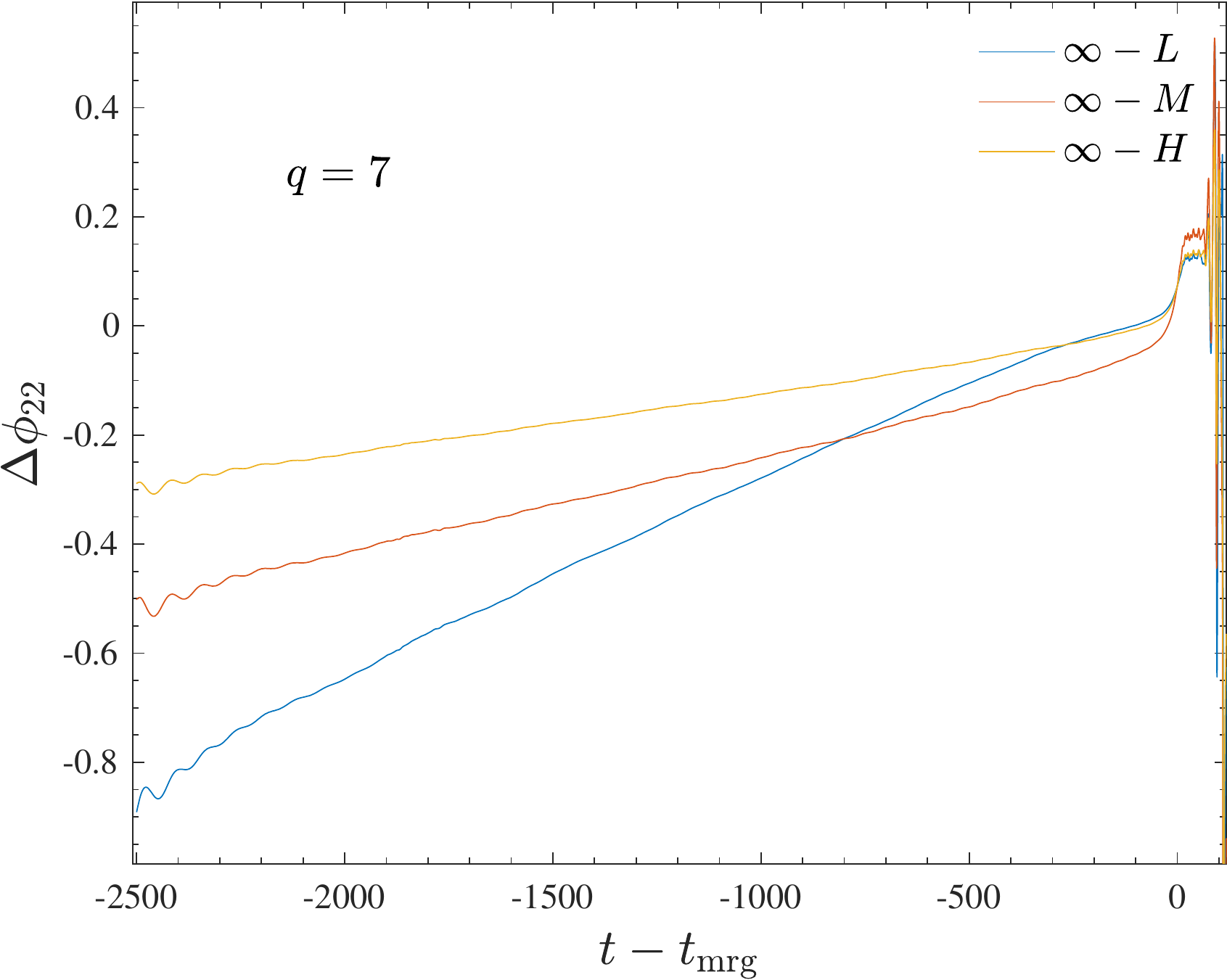}
\hspace{2mm}
\includegraphics[width=0.31\textwidth]{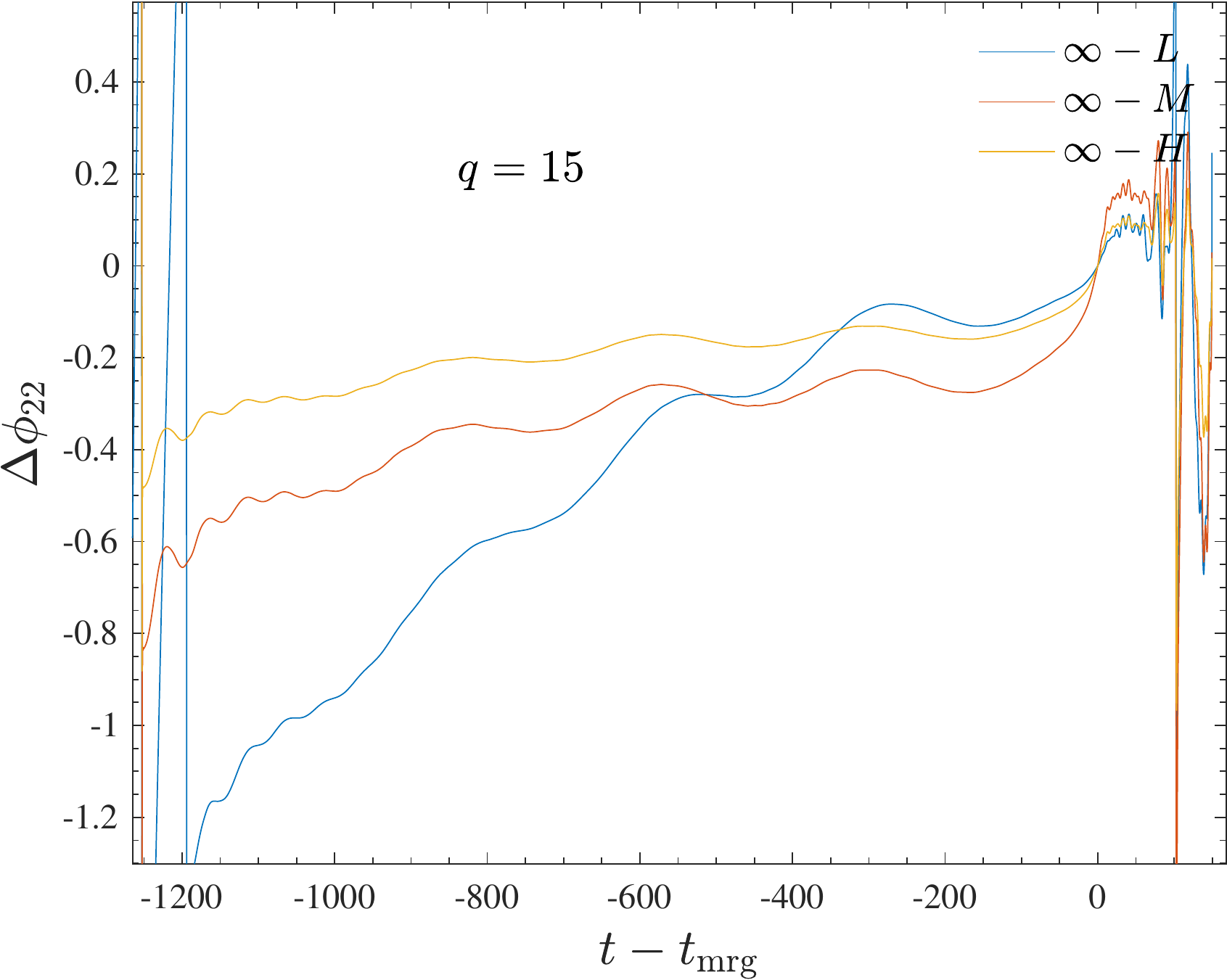}
\hspace{2mm}
\includegraphics[width=0.31\textwidth]{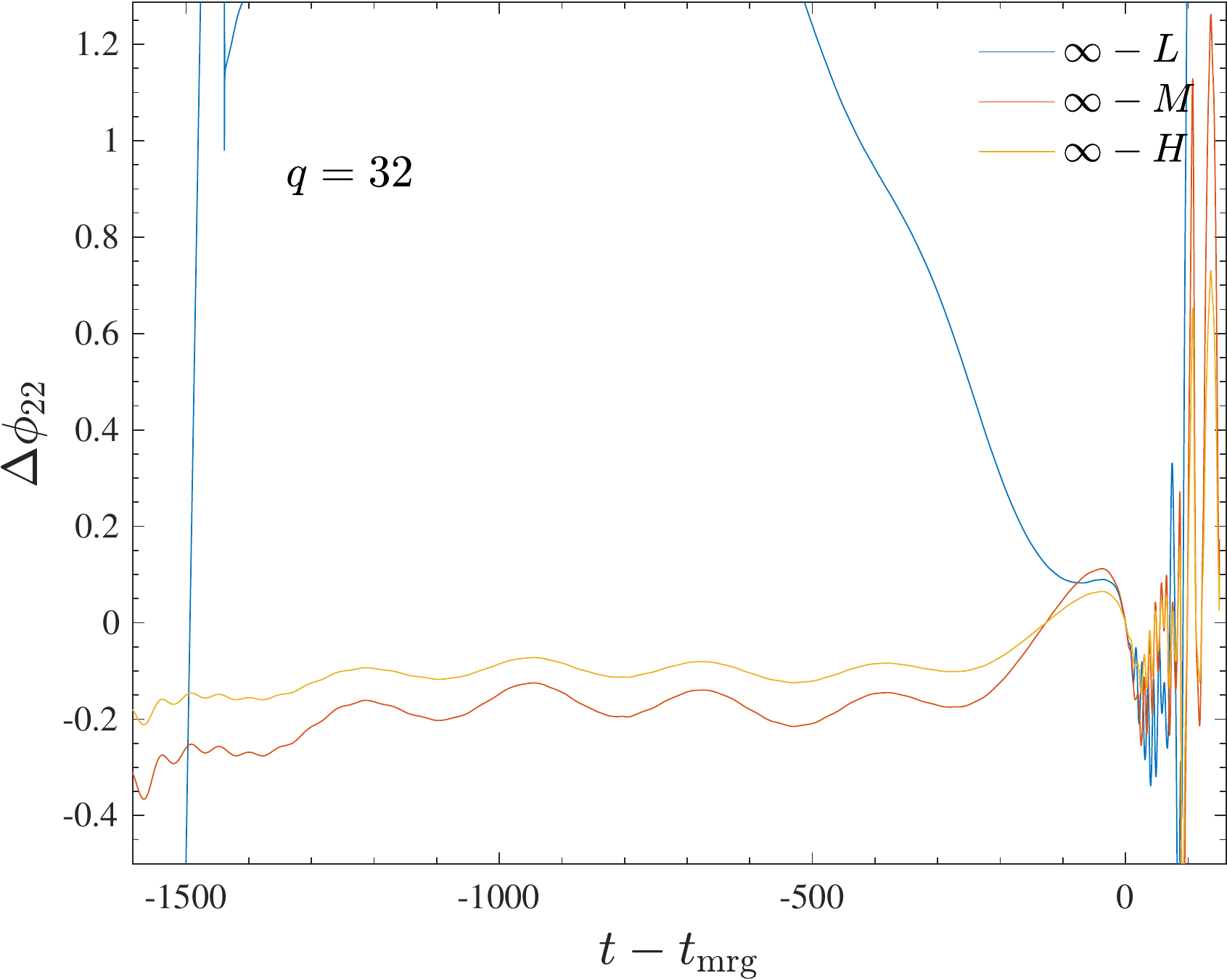}
\caption{\label{fig:rit_resolution}Estimate of the phase uncertainty on our simulations with $q=\{7,15,32\}$: 
accumulated phase differences between finite resolution waveform data and resolution extrapolated waveform 
(labelled with $\infty$). We consider three resolutions:Low (L), Medium (M), and High (H). The slope of the linear 
drift decreases as the resolution increases. The NR-RIT waveforms are aligned at merger time.}
\end{figure*}

Beside the direct use of NR simulations, the analysis of GW sources is mostly
done using waveform models that are obtained from the synergy between 
analytical and numerical relativity results.
The effective one body (EOB) approach~\cite{Buonanno:1998gg,Buonanno:2000ef,Damour:2000we,Damour:2001tu,Damour:2015isa} 
is a way to deal with the general-relativistic two-body problem that, by construction,
allows the inclusion of perturbative (e.g. obtained using post-Newtonian methods) 
and full numerical relativity (NR) results within a single theoretical framework. 
It currently represents a state-of-the-art approach for modeling dynamics and 
waveforms from binary black holes, conceptually designed to describe the entire
inspiral-merger-ringdown phenomenology of quasicircular
binaries~\cite{Nagar:2018gnk,Nagar:2018zoe,Cotesta:2018fcv,Nagar:2019wds,Nagar:2020pcj,Ossokine:2020kjp,Schmidt:2020yuu}
or even eccentric inspirals~\cite{Chiaramello:2020ehz,Nagar:2021gss,Nagar:2021xnh} 
and dynamical captures along hyperbolic orbits~\cite{Damour:2014afa,Nagar:2020xsk,Nagar:2021gss,Gamba:2021gap}.

The \teob{} model is the EOB waveform model that currently shows the highest level of NR-faithfulness~\cite{Albertini:2021tbt}
against all the spin-aligned NR waveforms available (see also Ref.~\cite{Akcay:2020qrj,Gamba:2021ydi} 
for the precessing case). The model has been tested~\cite{Nagar:2020pcj,Riemenschneider:2021ppj} 
against NR simulations available up to $q=18$. Although the model generates waveforms that look qualitatively
sane and robust also for larger mass ratios, only a direct comparison with NR data can effectively probe its performance 
in the large-$q$ regime.

The aim of this paper is to provide  EOB/NR waveform comparisons to validate the \teob{}
model (at least) up to $q=128$. To do so, we exploit the NR waveform data discussed above 
and presented in Ref.~\cite{Lousto:2020tnb}.
This paper is organized as follows:
Section~\ref{sec:eobnr} reviews both  the NR waveforms we are going to use
and the basics of the EOB model \teob{}. Section~\ref{sec:peak} exploits various sets of
NR data to probe the universal behavior of the multipolar hierarchy of waveform amplitudes
at merger, showing consistency with test-mass results. The EOB/NR phasing comparisons
are discussed in Sec.~\ref{sec:phasing}, while Sec.~\ref{sec:EOBNR} reports a few considerations
about the impact of NR systematics on informing EOB waveform models.
Concluding remarks are collected in Sec.~\ref{sec:end}. We use geometrized units with $c=G=1$.

\section{NR and EOB waveform data}
\label{sec:eobnr}
Let us start by fixing our waveform conventions. The multipolar decomposition
of the strain waveform is given by
\be
\label{eq:h_pm}
h_+-ih_\times = {\cal D}_L^{-1}\sum_{\lm} h_{\lm} {}_{-2}Y_{\lm}(\theta,\varphi),
\ee
where ${\cal D}_L$ is the luminosity distance and  ${}_{-2}Y_{\lm}(\theta,\varphi)$ are 
the $s=-2$ spin-weighted spherical harmonics. For consistency with previous works
involving the \teob{} waveform model, we work with Regge-Wheeler-Zerilli normalized
multipoles~\cite{Nagar:2005ea,Nagar:2006xv} defined as $\Psi_\lm = h_\lm/\sqrt{(\ell+2)(\ell+1)\ell(\ell-1)}$,
and each mode is decomposed in amplitude and phase
\be
\Psi_\lm = A_\lm\, e^{-i\phi_\lm} \ .
\ee
The binary has masses $(m_1,m_2)$. We adopt the convention that $m_1\geq m_2$ and thus we
define $q\equiv m_1/m_2\geq 1$, $M=m_1+m_2$ and the symmetric mass ratio as $\nu\equiv m_1 m_2/M$.

\subsection{NR simulations}
\label{sec:nr}
We use here the simulations presented in Ref.~\cite{Lousto:2020tnb}, to which we refer the reader
for additional technical details (see also Refs.~\cite{Healy:2020iuc,Rosato:2021jsq}). 
Reference~\cite{Healy:2020iuc} explored different gauge choices in the moving puncture formulation
in order to improve the accuracy of a linear momentum measure evaluated
on the horizon of the remnant black hole produced by the merger of a binary.
Similarly, Ref.~\cite{Rosato:2021jsq} investigated the benefits of adapted gauges 
to large mass ratio binary black hole evolutions. 
We found expressions that approximate the late time behavior of the lapse and shift, 
$(\alpha_0,\beta_0)$, and use a position and black hole
mass dependent shift damping term, $\eta[\vec{x}_1(t),\vec{x}_2(t),m_1,m_2]$.
We found that this substantially
reduces noise generation at the start of the numerical integration
and keeps the numerical grid stable around both black holes, allowing for more accuracy with
lower resolutions. We tested this gauge in detail in a case study of a binary with a
7:1 mass ratio, and then use 15:1 and 32:1 binaries for a convergence study. 
NR waveforms~\cite{Abbott:2016apu,Lange:2017wki} 
are being directly applied to GW parameter estimation,
demonstrating  how source parameters
for generic BHBs can be inferred based directly on solutions of Einstein's equations. 
Specific cases have been performed for the GW150914 and GW170104~\cite{Lovelace:2016uwp,Healy:2017xwx,Kumar:2018hml} events,
finding excellent agreement between RIT and 
SXS~\cite{Buchman:2012dw,Chu:2009md,Hemberger:2013hsa,Scheel:2014ina,Blackman:2015pia,Lovelace:2011nu,
Lovelace:2010ne,Mroue:2013xna,Lovelace:2014twa,Kumar:2015tha,Lovelace:2016uwp,Abbott:2016nmj,Boyle:2019kee} 
waveforms up to $\ell=m=5$ modes,
but for comparable masses between $q=1/0.85$ and $q=1/0.43$.
The direct use of theoretical waveform information to interpret
gravitational waves
observations and to determine the precise nature of the astrophysical
sources has proven to be a remarkable success when applied to O1/O2
BBH events~\cite{Healy:2020jjs} and beyond~\cite{Gayathri:2020fbl}.
And the recent release of the RIT binary black hole waveform public catalog
includes 1881 simulations~\cite{Healy:2022wdn}.

We only report
here the information that is pertinent for our targeted EOB/NR comparisons. In particular,
let us remember that we follow Ref.~\cite{Healy:2017zqj} to setup quasi-circular initial data 
that allow our simulations to have a negligible amount of spurious initial junk radiation. 
Similarly, we use the procedure of Ref.~\cite{Nakano:2015pta} to accurately extrapolate the
waveform to infinity. The code natively outputs the Weyl scalar $\Psi_4$ that is then transformed 
to the strain by applying a standard integration procedure in the frequency domain~\cite{Reisswig:2010di}.
We consider mass ratios $q=(7,15,32,64,128)$. For $q=(7,15,32)$ we could complete runs 
at three resolutions, named Low (L), Medium (M) and High (H), so to have a complete convergent series. 
This allowed us to Richardson-extrapolate 
the waveform to infinite resolution and thus give an estimate of the phase uncertainty.
Figure~\ref{fig:rit_resolution} reports the phase differences between the resolution-extrapolated $\ell=m=2$ 
waveform (indicated as $\infty$) and each finite resolution. All waveforms are aligned at merger 
point (marked with $t-t_{\rm mrg}=0$), where merger is defined as the peak of the quadrupolar 
amplitude $A_{22}$.
We remark that the $\Delta\phi_{22}(t)\equiv \phi^{\infty}_{22}-\phi_{22}^{L,M,H}$ 
is fundamentally a linear function of time during the inspiral up to merger, and its slope {\it decreases} 
as the resolution is increased\footnote{Note however the different behavior for $q=32$, suggesting 
that the low resolution is not high enough to correctly capture this behavior.}.
This, well known, effect related to resolution will be useful later when interpreting EOB/NR phase comparison.
Finally, Fig.~\ref{fig:rit_resolution} indicates that a phase uncertainty $\Delta\phi_{22}$ between 
$\sim 0.2$ and $\sim 0.4$~rad looks like a reasonable (conservative) error bar estimate on resolution 
extrapolated waveforms.

\subsection{Effective-one-body framework}
\label{sec:eob}
We use here the most advanced quasi-circular version of  the 
\teob{} model~\cite{Nagar:2020pcj,Riemenschneider:2021ppj}, 
that also includes several subdominant modes completed through 
merger and ringdown. More precisely,
we use here the {\tt Matlab} private implementation of the model
(and not the public one written in $C$~\cite{Riemenschneider:2021ppj})
that relies on the iterative determination of the next-to-quasi-circular
correction parameters and not on the fits described in Ref.~\cite{Riemenschneider:2021ppj}.
Let us recall that this waveform model exploits NR waveform data in
two ways. On the one hand, NR waveforms are used to inform two
dynamical parameters that enter directly the EOB Hamiltonian 
(both in the orbital and spin-orbital sector). On the other hand, NR waveform data 
are used in the description of merger and ringdown via a certain fitting 
procedure~\cite{Damour:2014yha}  of NR data. The model exploited SXS data
up to $q=10$, with five more datasets with $q=18$~\cite{Nagar:2020pcj}. 
Test-mass data obtained using {\tt Teukode}~\cite{Harms:2014dqa} are 
also used to inform the fits  of amplitude and frequency at peak of each multipole.
The resulting \teob{} waveform model has been validated against several hundreds of
NR simulations, of different accuracy, including mass ratios up to $q=18$. 
For larger mass ratios the model generates waveforms that looks sane, in general
non pathological (excluding extremely spinning cases), but a systematic validation
of their quality has not been done so far for the lack of suitable NR data, 
and it will be the focus of Sec.~\ref{sec:phasing} below.

\section{Multipolar hierarchy at merger}
\label{sec:peak}

\begin{figure*}[t]
	\center
\includegraphics[width=0.22\textwidth]{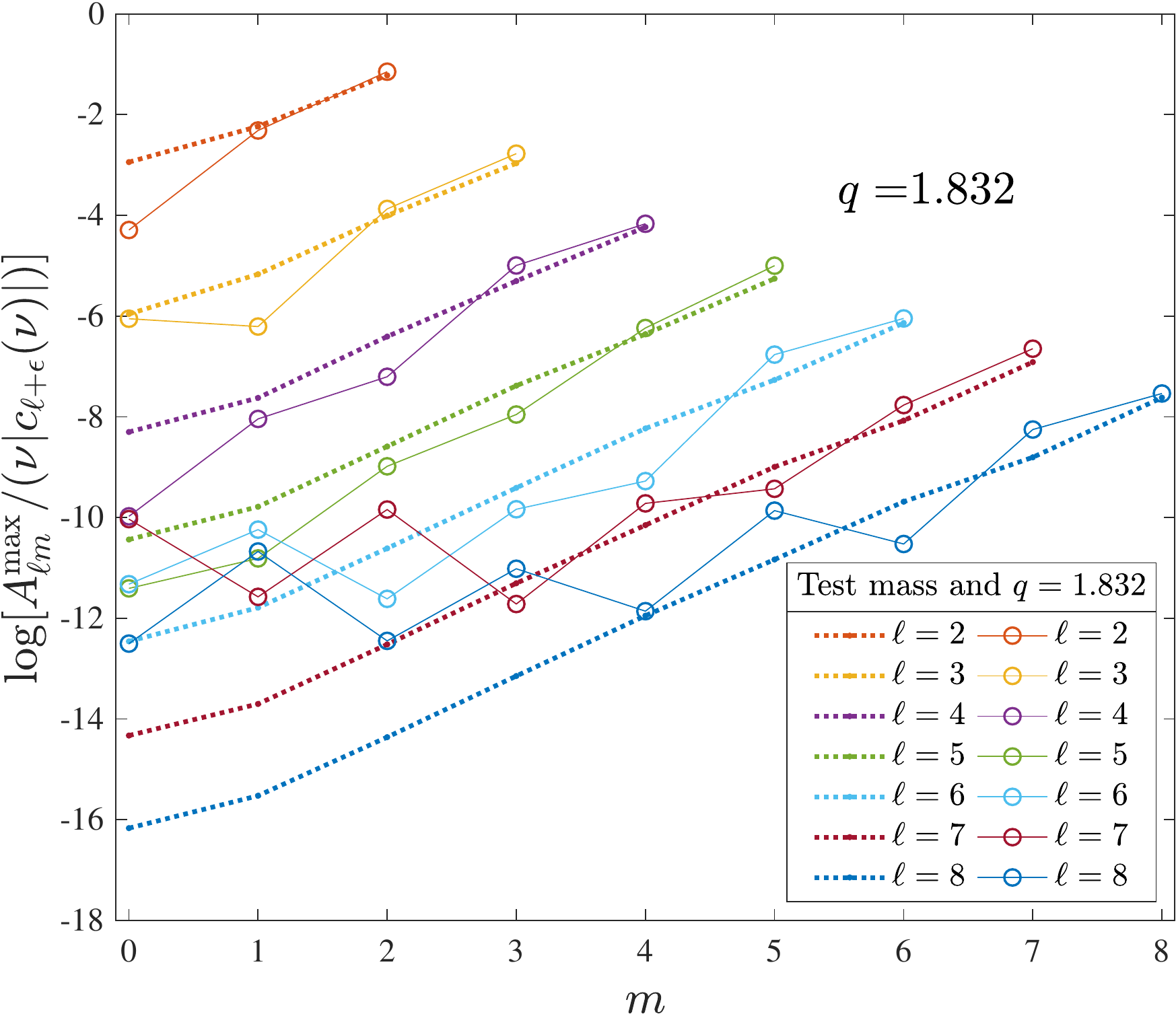}	
\includegraphics[width=0.22\textwidth]{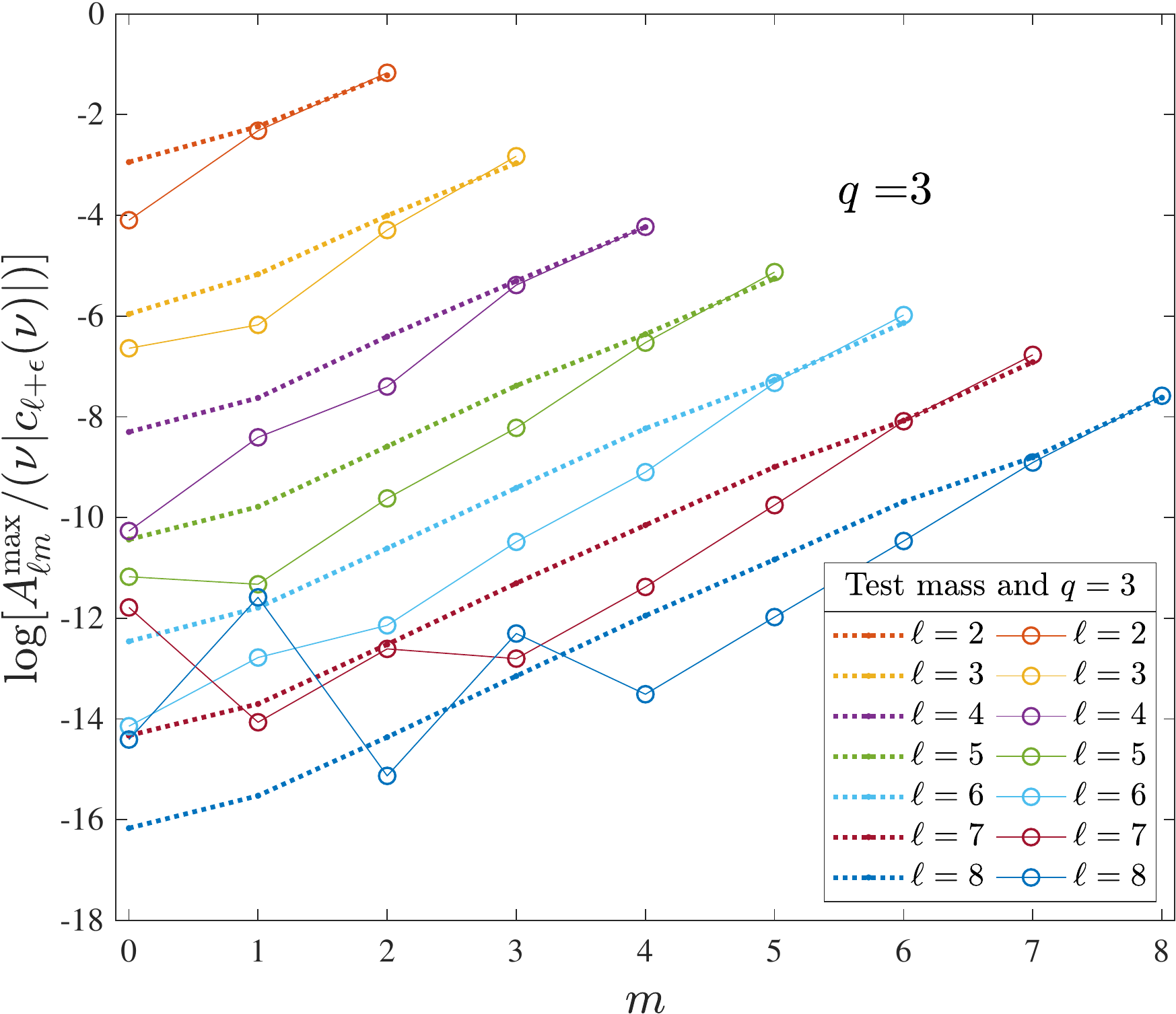} 
\includegraphics[width=0.22\textwidth]{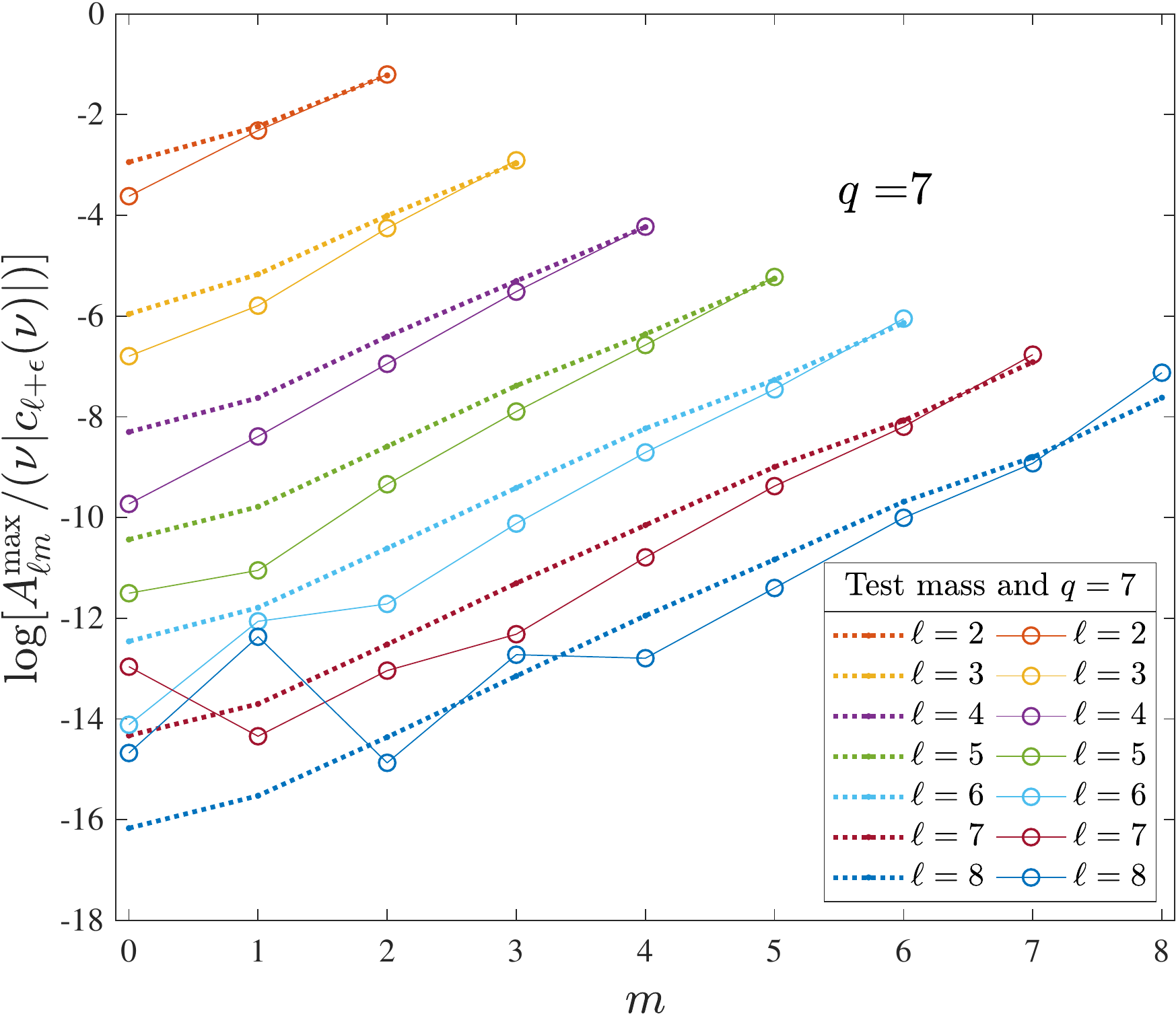} 
\includegraphics[width=0.22\textwidth]{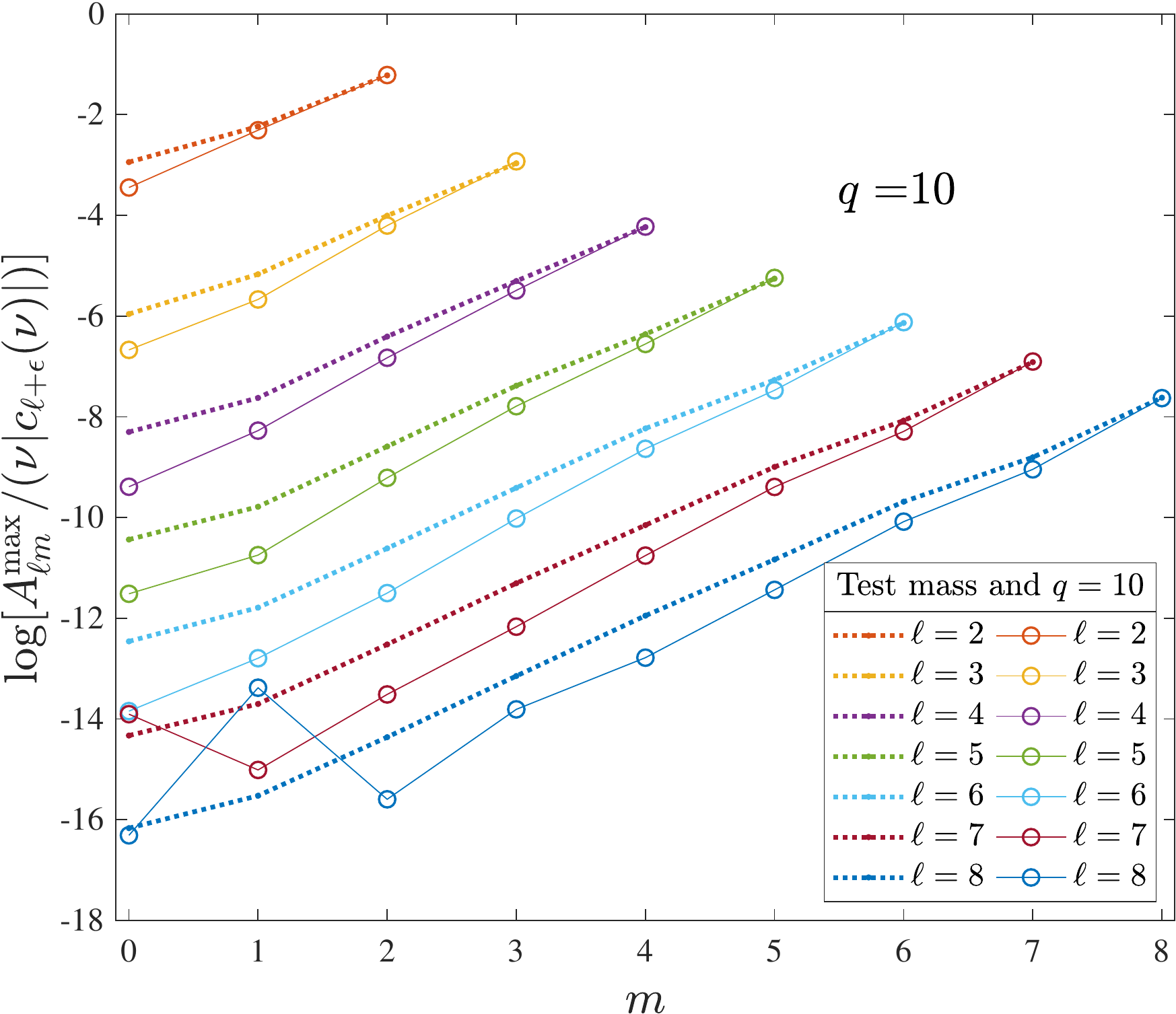} \\ 
\includegraphics[width=0.22\textwidth]{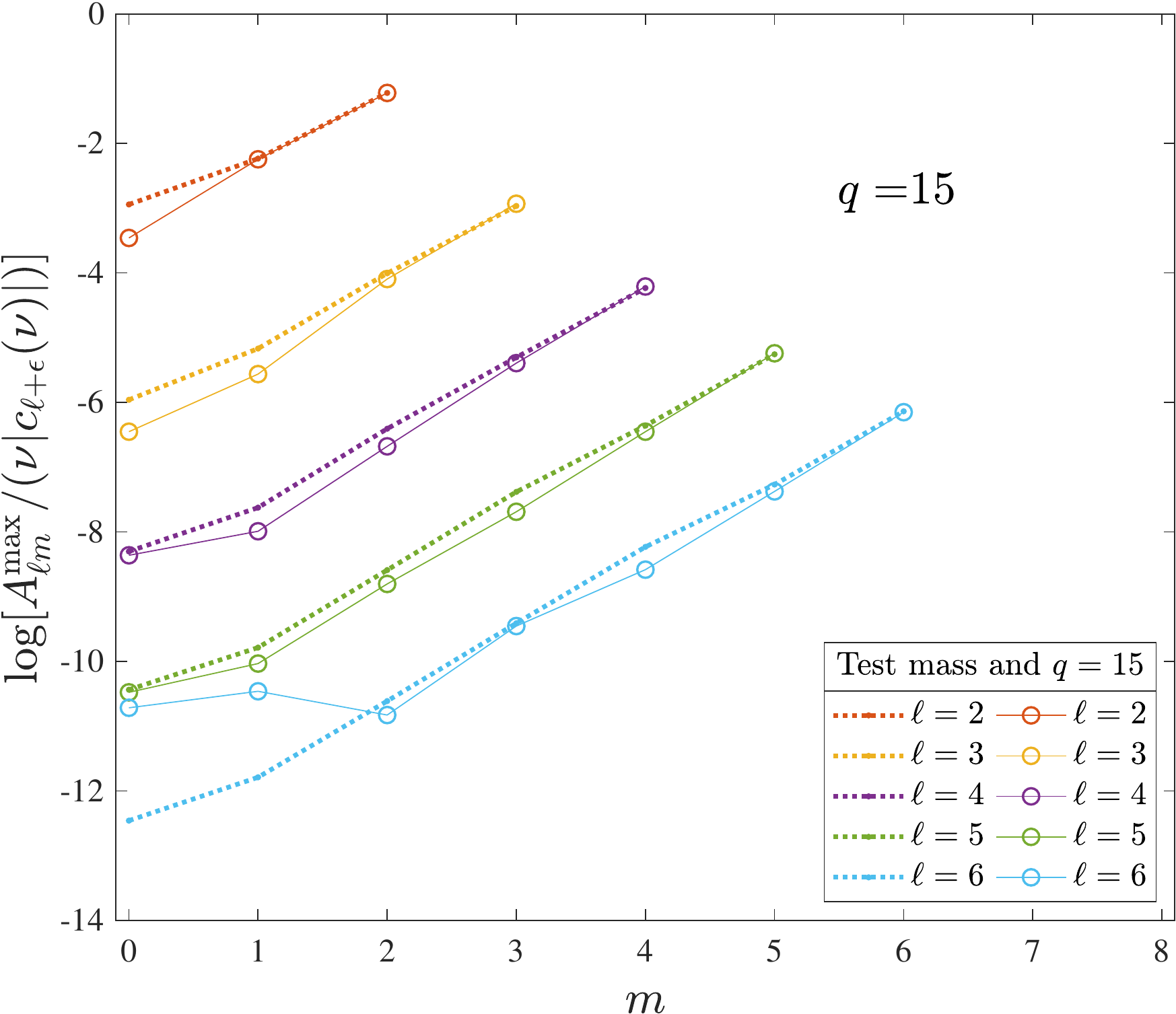}	
\includegraphics[width=0.22\textwidth]{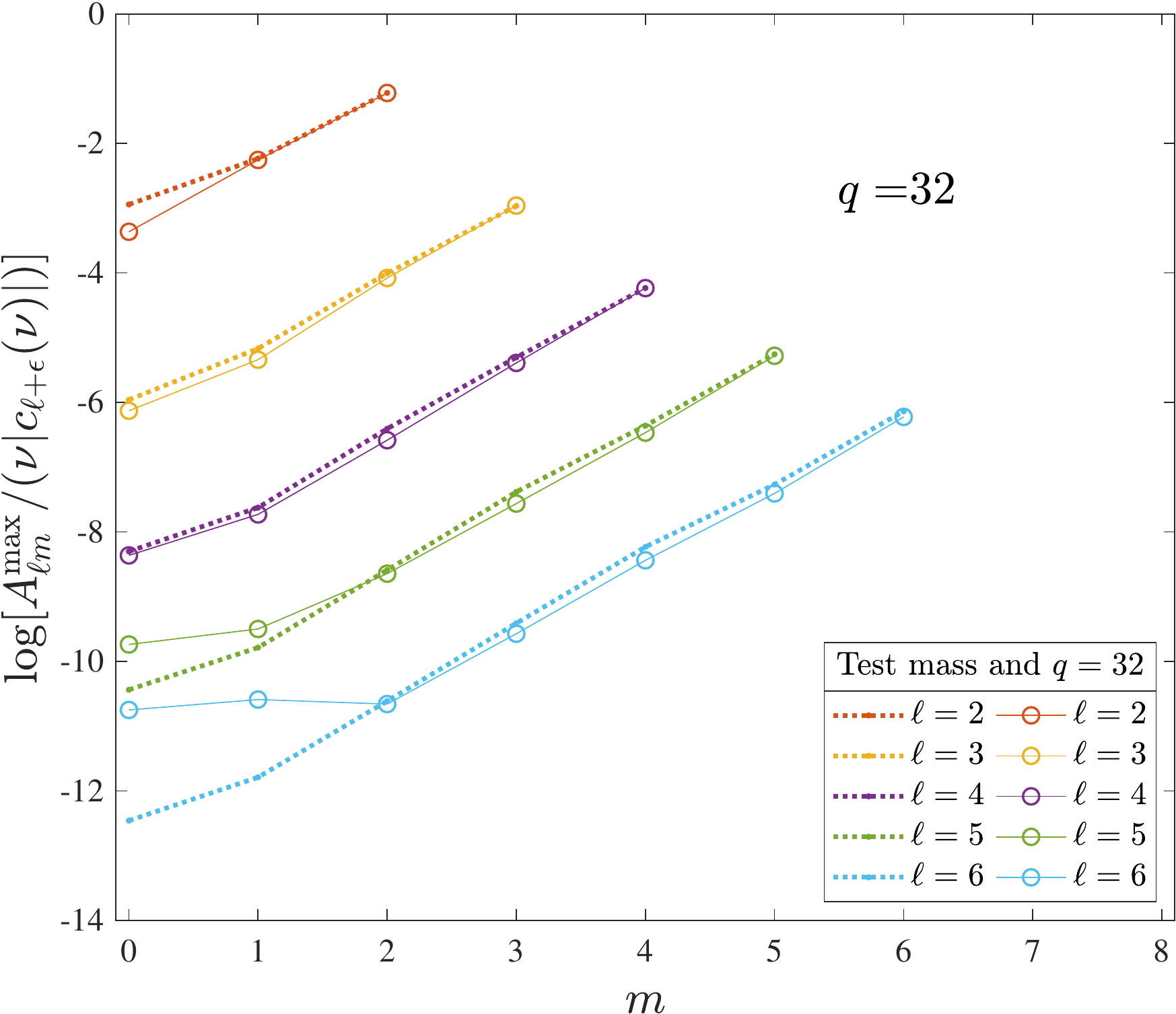}
\includegraphics[width=0.22\textwidth]{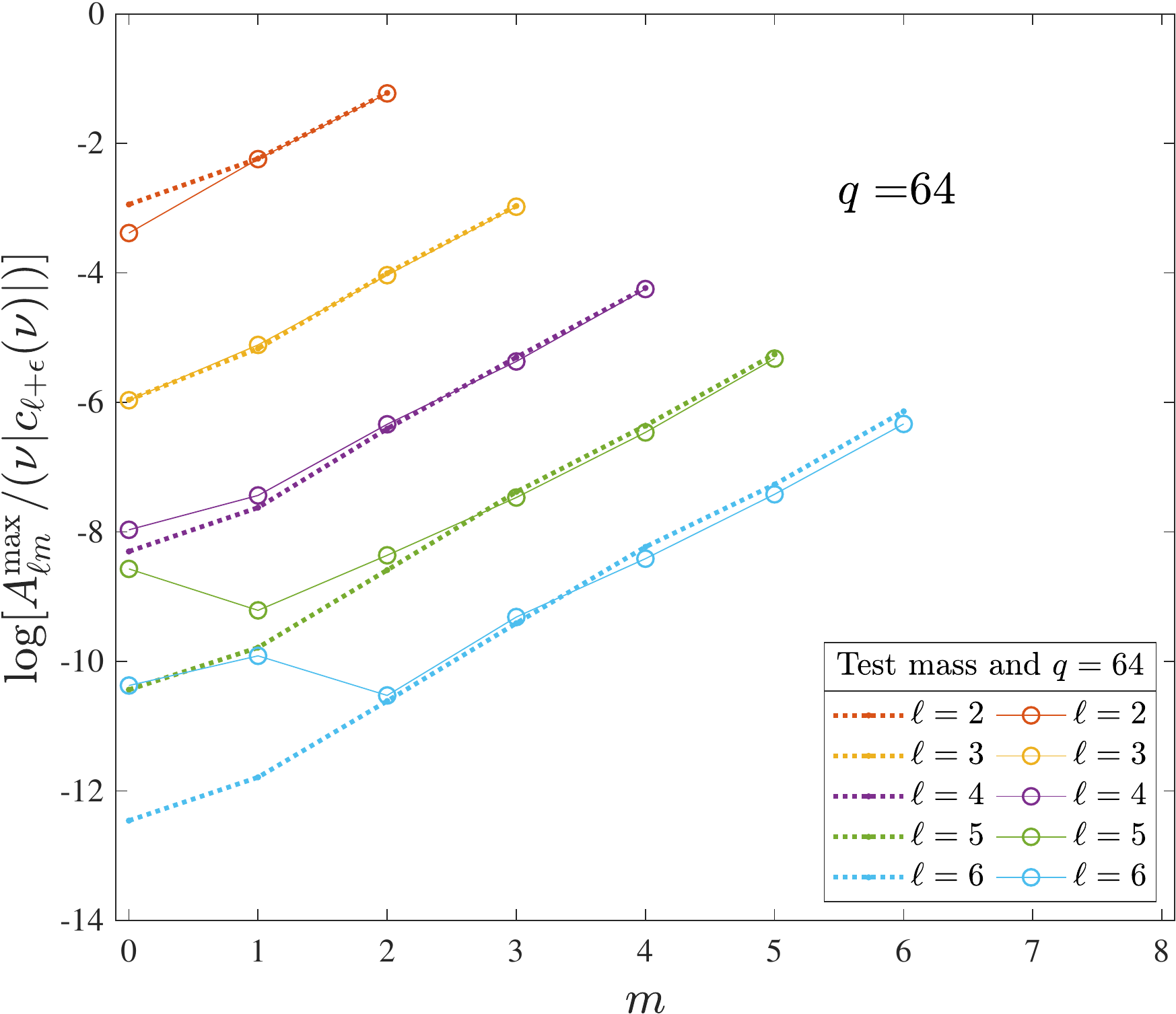}
\includegraphics[width=0.22\textwidth]{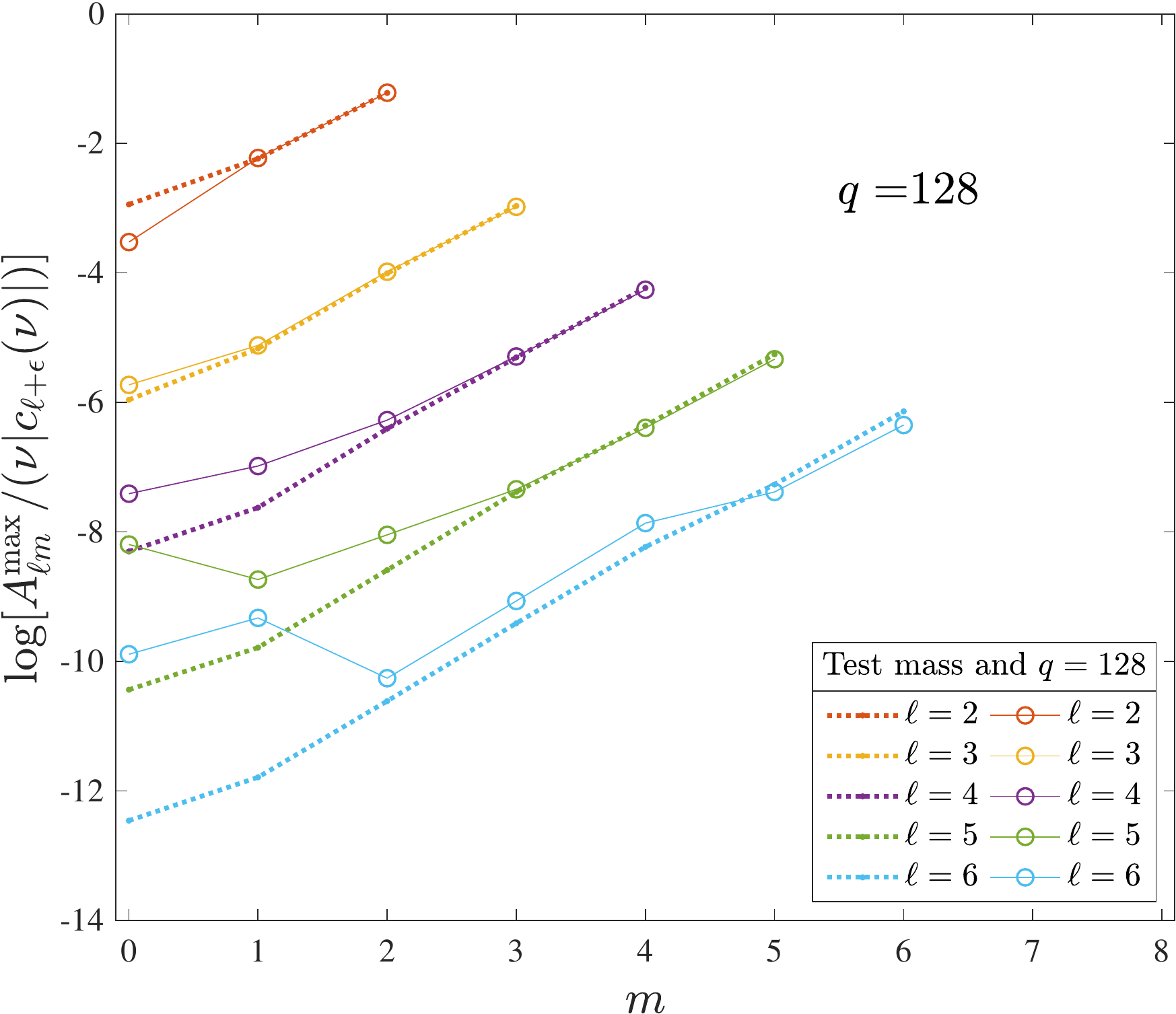}
\caption{\label{fig:A_mrg}Multipolar hierarchy of  merger amplitudes: $\nu$-reduced maximum amplitude for each multipole,
Eq.~\eqref{eq:Alm_max_nu} versus $\ell$ and $m$. The test-mass values (dotted lines) are compared with the NR values for
various $q$. Up to $q=10$ we use SXS data. For $q>10$
(bottom panels) we use only RIT waveforms. 
Note the consistency 
between all the $\ell=m$ modes. The plot highlights the well-known decrease of importance of the subdominant multipoles 
with $m<\ell$ as the mass ratio is decreased. For each value of $\ell$, an approximate exponential dependence on $m$ is found, 
with qualitative consistency between the test-mass and the comparable-mass cases. The oscillations present in the SXS $\ell=7$ and 
$\ell=8$ modes for small values of $m$ denote inaccuracies in the simulations.}
\end{figure*}

\begin{figure*}[t]
	\center
\includegraphics[width=0.31\textwidth]{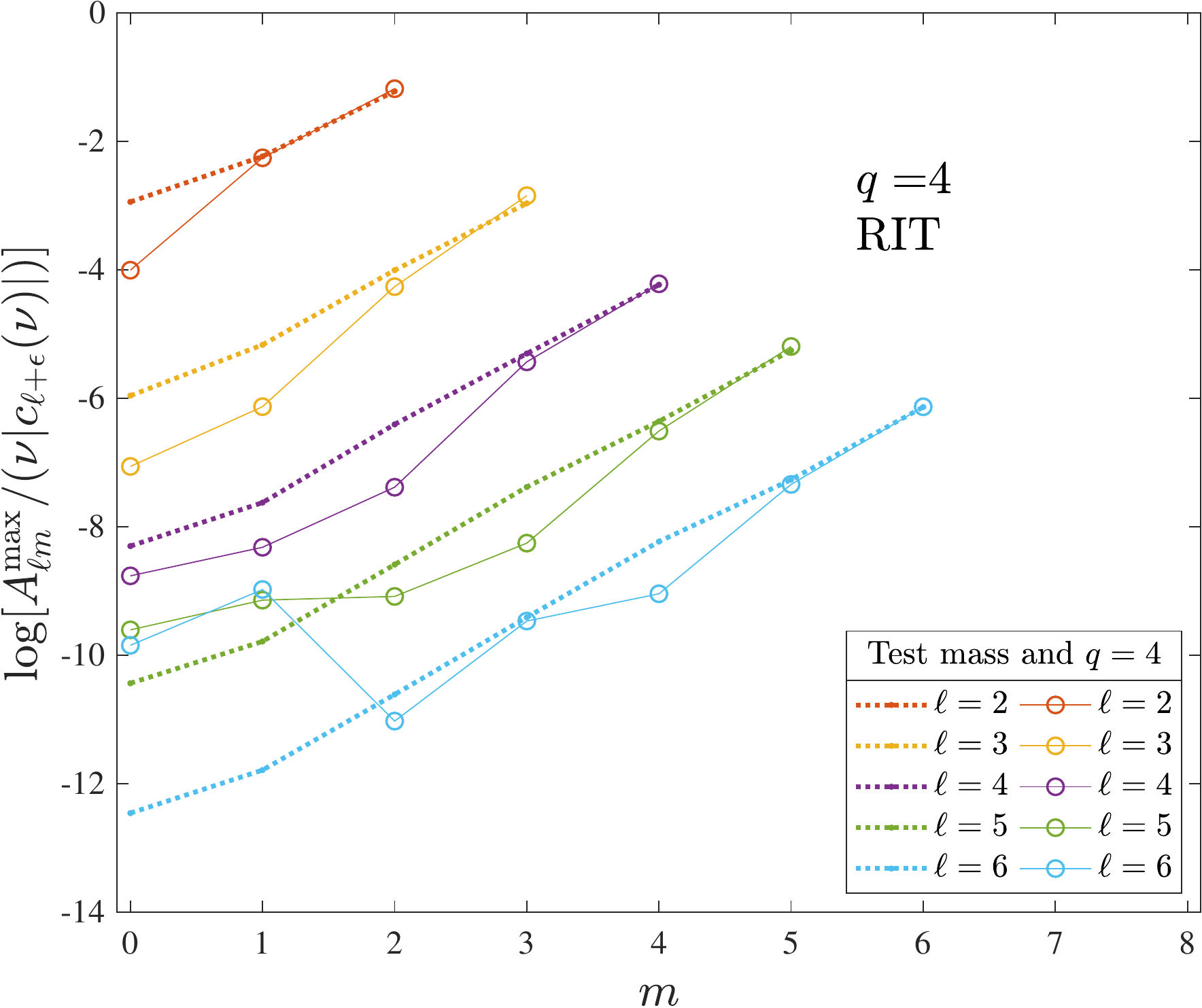} 
\hspace{2mm}
\includegraphics[width=0.31\textwidth]{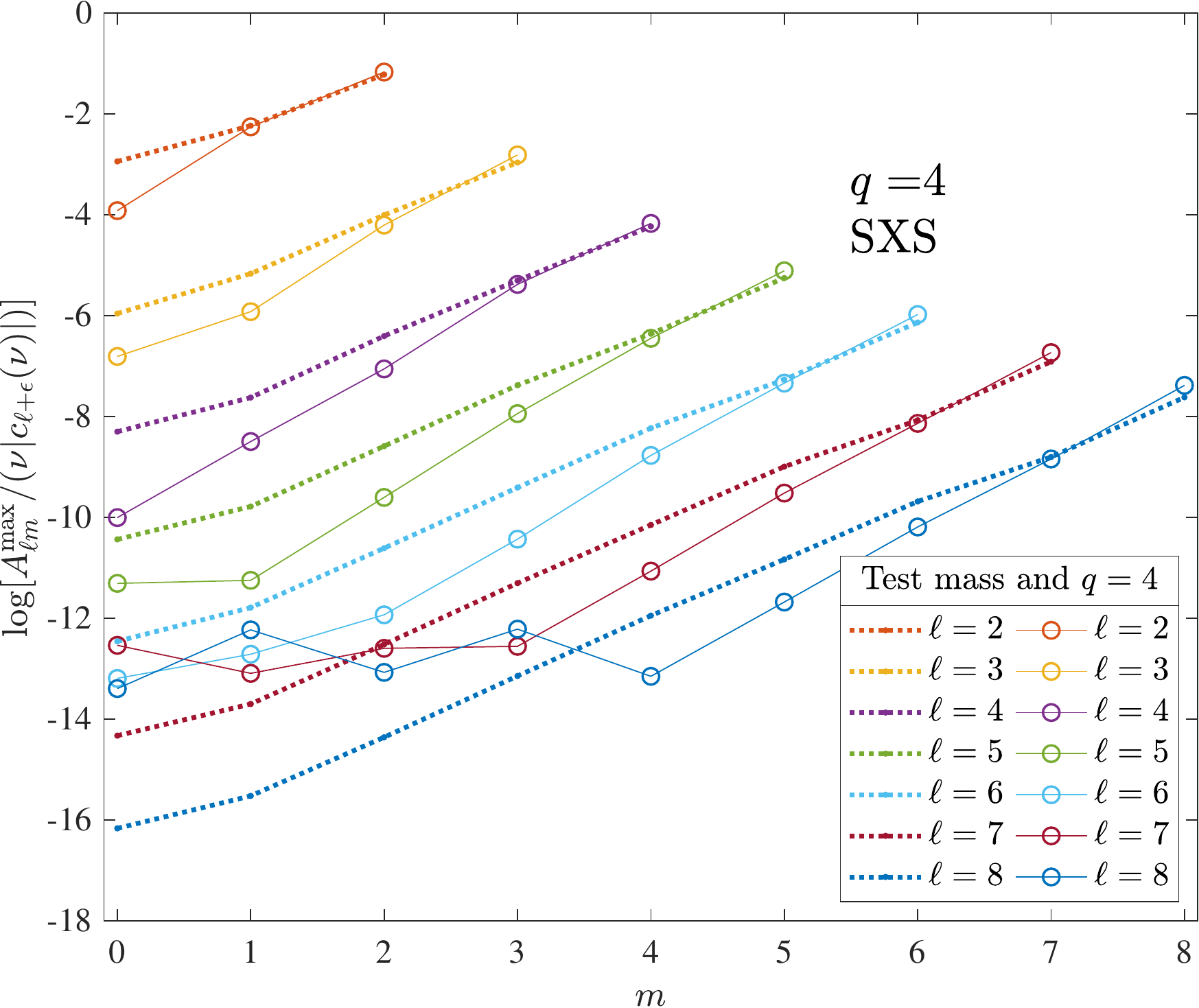}	
\hspace{2mm}
\includegraphics[width=0.31\textwidth]{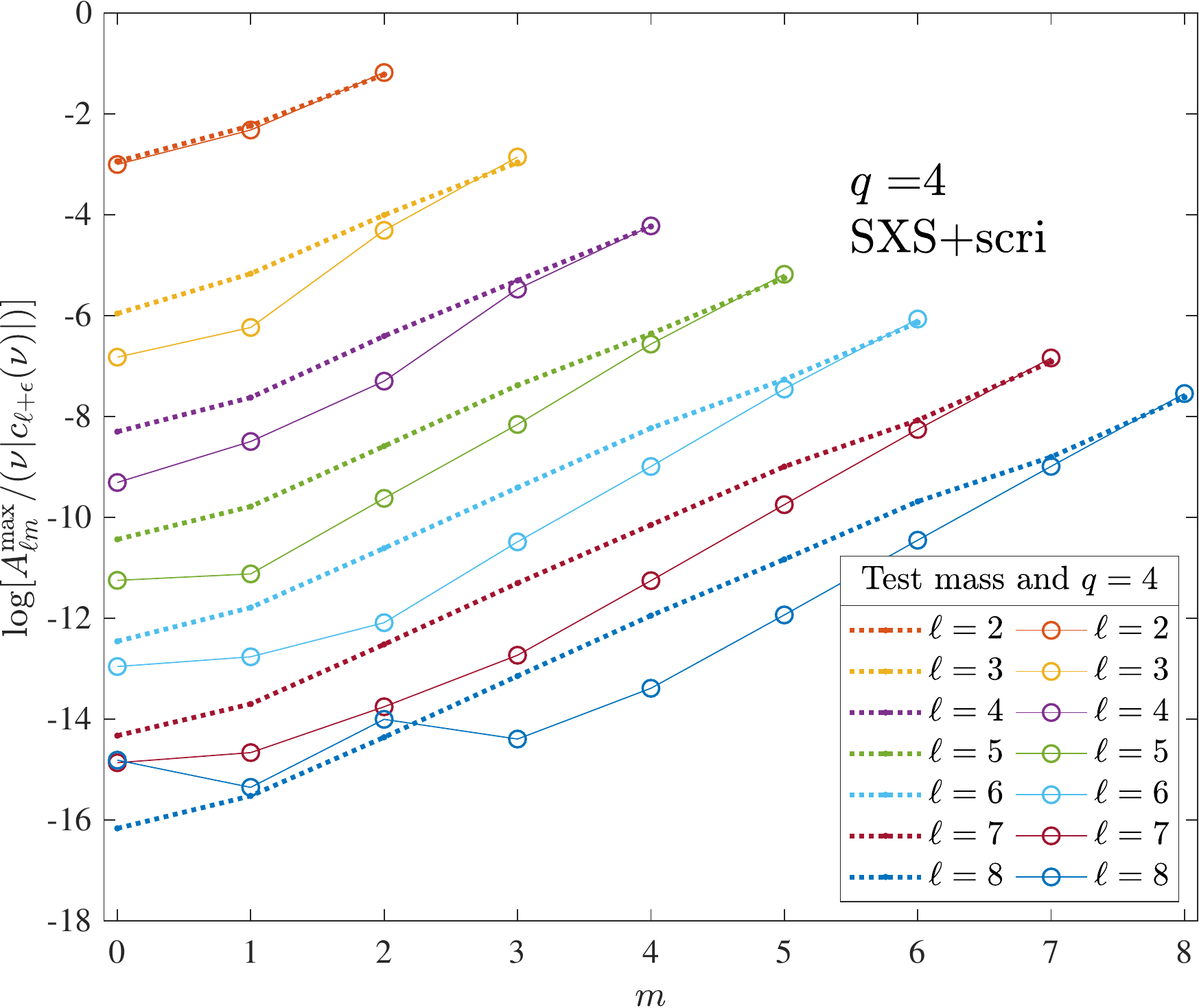} 
\caption{\label{fig:A_mrg_spectre}Multipolar hierarchy of  merger amplitudes for $q=4$ from
  RIT data (left panel), 
  standard SXS data (central panel, SXS:BBH:2030 dataset), and 
  SXS waveforms extrapolated to $\mathscr{I}^+$ using the python 
package \texttt{scri} of Ref.~\cite{Iozzo:2020jcu} (right panel) with the addition of the contribution of the displacement memory as
described in Ref.~\cite{Mitman:2020bjf}.
Note the improved consistency of the \texttt{scri}-extrapolated data and test-mass data for all $\ell=2$ multipoles 
as well as for the $\ell=m$ ones.}
\end{figure*}

The multipolar structure of the waveform amplitude at merger has a universal structure that can be 
described by the mass ratio $\nu$ and an effective spin parameter. This hierarchy emerges when 
leading order PN expressions are suitably factorized. In particular, Ref.~\cite{Nagar:2018zoe}, pointed 
out a quasi-universal behavior in 
the symmetric mass ratio $\nu$ and spin parameter $\hat{S}$ of the $\ell=m=2$ merger frequency (see Fig.~33 therein). 
Working in the test-mass limit, Ref.~\cite{Bernuzzi:2010xj} identified a simple structure of the 
multipolar peak amplitudes $A_\lm^{\rm max}=\max_t(A_\lm)$ in terms of the multipolar order $(\ell,m)$, 
\be
\label{eq:Alm_max}
A_\lm^{\rm max}/\nu\approx e^{c_1(\ell) m + c_2(\ell)\ell},
\ee
where the coefficients $c_i(\ell)$ are listed in Table~VI of~\cite{Bernuzzi:2010xj}; for $\ell>2$, 
the coefficient $c_1$ is practically independent of $\ell$.
Here we show that this structure is present in any BBH multipolar waveform and can be recovered 
by analytically removing the leading $\nu$ dependence in each multipole.
On a practical level, this finding is important to construct the ringdown part of the EOB 
waveform~\cite{Damour:2014yha} in order to design accurate and physically motivated fits to NR data.

From PN theory (see e.g.~\cite{Damour:2008gu}), the leading $\nu$ dependence of $A_\lm^{\rm max}$ is
\be
\label{eq:Alm_max_nu}
\doublehat{A}_\lm\equiv \dfrac{A_\lm^{\rm max}}{\nu |c_{\ell +\epsilon}(\nu)|}\,,
\ee
where
\be
c_{\ell + \epsilon}(\nu) = X_2^{\ell + \epsilon-1}+(-)^m X_1^{\ell + \epsilon-1}\,.
\ee
In the expressions above, $\epsilon\equiv \pi(\ell +m)$ is the parity of $\ell+m$, $\epsilon=0$ if $\ell+m$ is even
and $\epsilon=1$ if $\ell+m$ is odd. Although this structure is used in Refs.~\cite{Nagar:2018zoe,Nagar:2021gss} 
to accurately fit the  multipolar amplitude values around merger, it has not been spelled out 
explicitly before, and in particular not in connection with test-mass results.
Figure~\ref{fig:A_mrg} contrasts the values of $\doublehat{A}_\lm$ for several  comparable 
mass binaries (solid lines) with the corresponding test-mass values taken from Ref.~\cite{Bernuzzi:2010xj}
(dotted lines) up to $\ell_{\rm max}=8$, when available. We use the following SXS datasets:
SXS:BBH:1354 ($q=1.832$); SXS:BBH:1178 ($q=3$); SXS:BBH:0298 ($q=7$) and SXS:BBH:1107 
($q=10$). Each dataset is taken at the highest resolution available and choosing $N=2$ extrapolation order
\footnote{Each waveform in the SXS catalog is available with extrapolation orders $N = 2, 3, 4$. The
general rule is to use $N=2$ when one is mostly interested in the late part of the waveform, choose $N =4$
for the inspiral and $N = 3$ for a compromise.
}~\cite{Boyle:2009vi}, in order to assure a more robust representation
of merger and ringdown part. 
For SXS data we have all multipoles up to $\ell_{\rm max}=8$, while our RIT waveforms are limited 
to $\ell_{\rm max}=6$ and we only focus on the $m=\ell$ and $m=\ell-1$ modes.
From Fig.~\ref{fig:A_mrg} one sees that the test-mass hierarchy between the modes is preserved
also in the comparable-mass case. The figure also highlights the {\it quantitative} consistency 
between the test-mass and comparable-mass $m=\ell$ and $m=\ell-1$ values of $\doublehat{A}_\lm$.
We observe a degradation of the accuracy of NR simulations with both low
values of $m$ and levels of radiation (high $\ell$).

\begin{figure*}[t]
	\center
\includegraphics[width=0.41\textwidth]{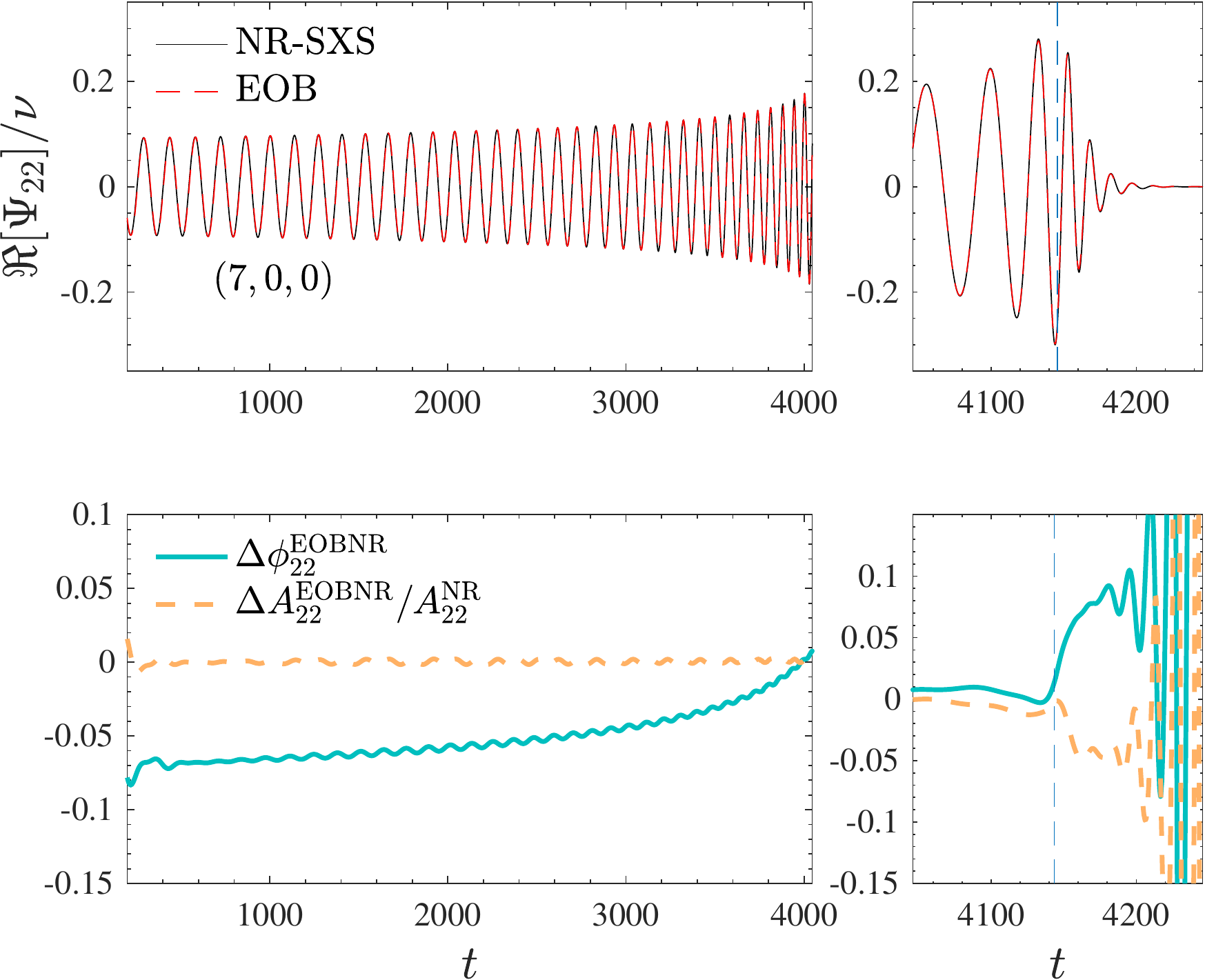}
\hspace{5mm}
\includegraphics[width=0.41\textwidth]{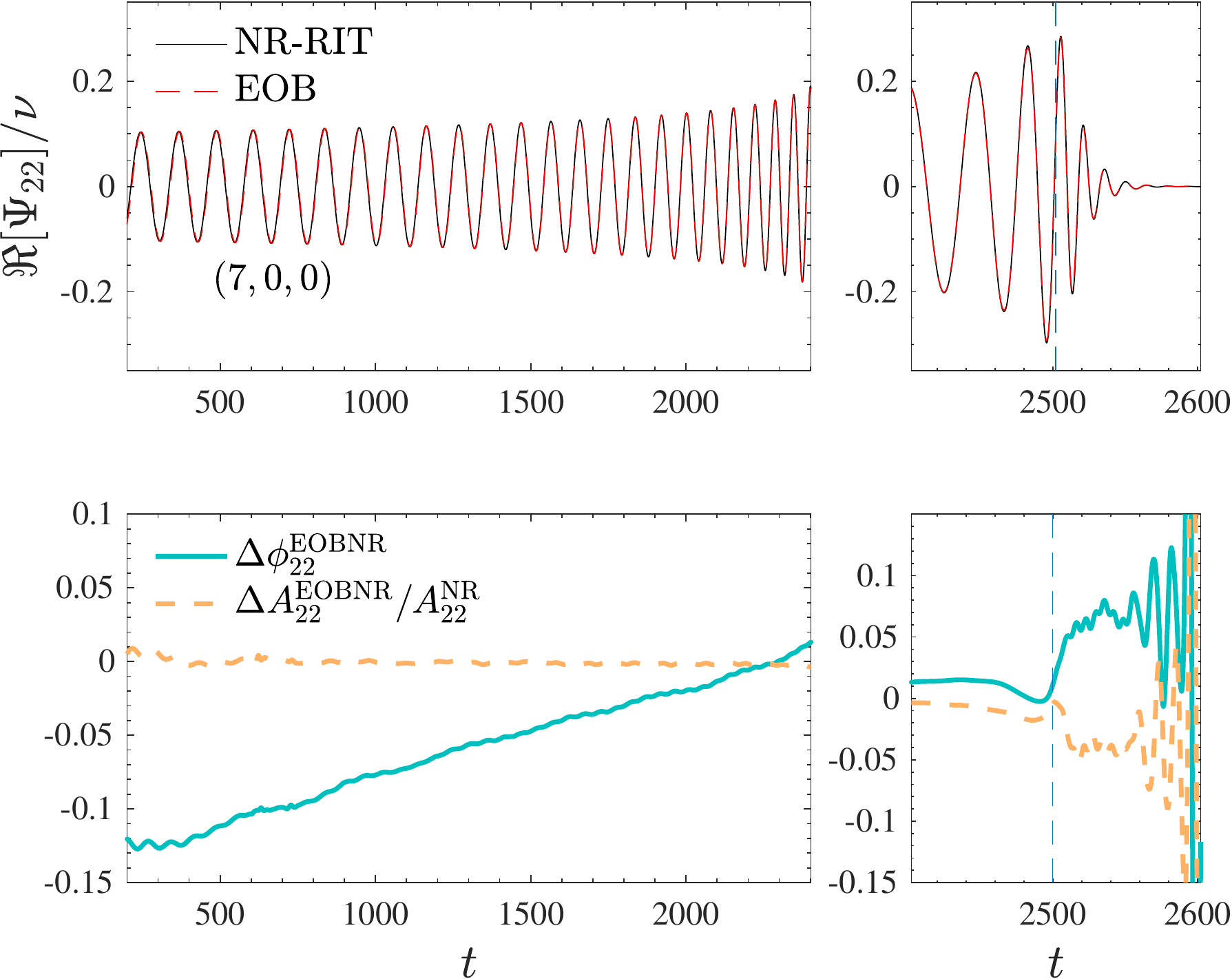}\\
\vspace{5mm}
\includegraphics[width=0.41\textwidth]{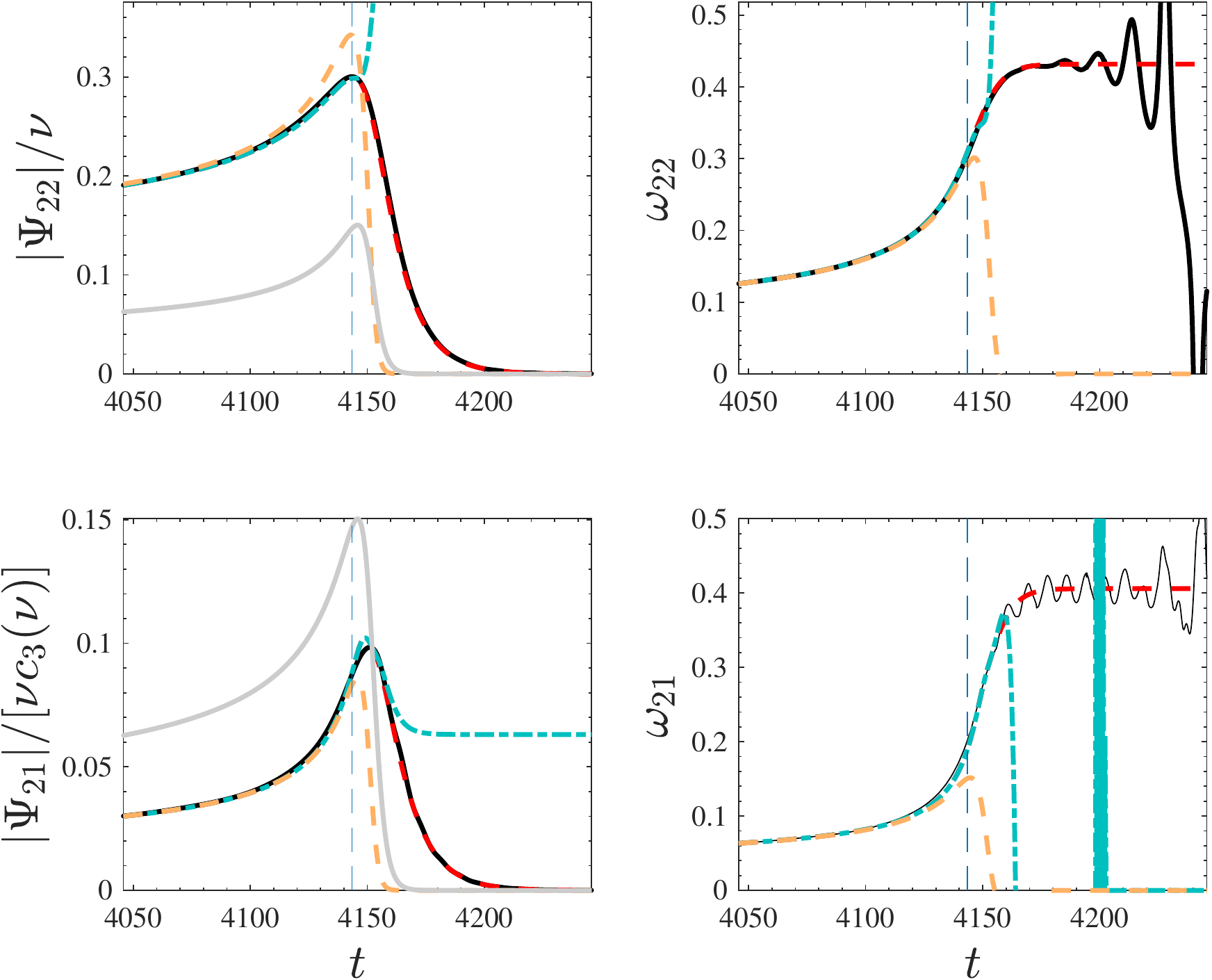}
\hspace{5mm}
\includegraphics[width=0.41\textwidth]{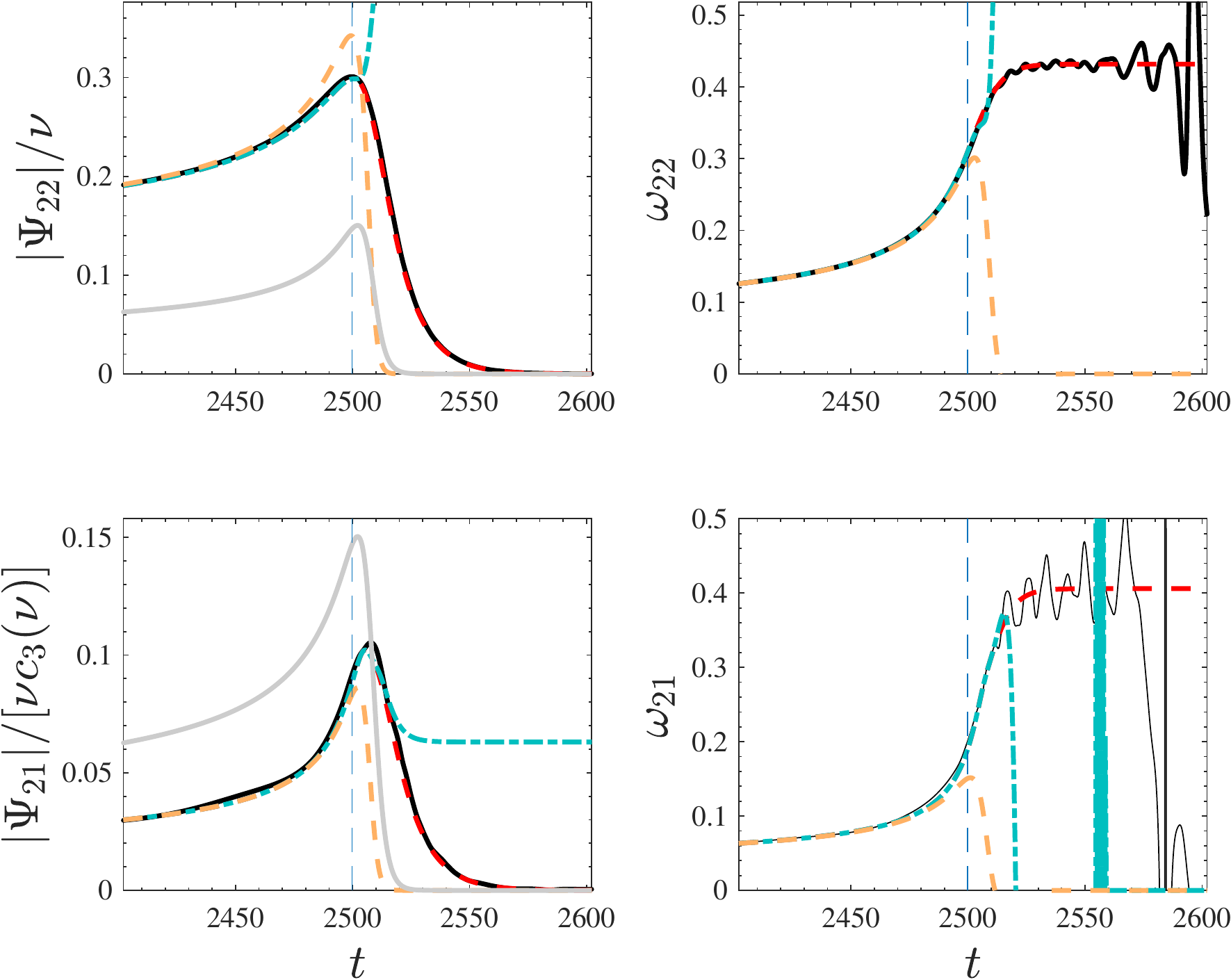}
\caption{\label{fig:sxs_vs_rit} EOB/NR comparison for $q=7$ configuration using either SXS:BBH:0298 (left column) or 
the resolution-extrapolated RIT data presented here (right column). Waveform are aligned on the frequency interval $[\omega_L,\omega_R]=[0.2,0.3]$,
close to merger. The top row report the real part of the $\ell=m=2$ mode as well as the phase difference, 
$\Delta\phi_{22}^{\rm EOBNR}\equiv \phi^{\rm EOB}_{22}-\phi^{\rm NR}_{22}$ 
and the relative amplitude difference. The bottom panels compare amplitude and frequencies of both the $(2,2)$ and $(2,1)$ waveform mode. 
The consistency between the two NR waveforms is remarkable, although the phase accumulated by the RIT one is globally slightly larger,
but coherent with the NR uncertainty estimate of Fig.~\ref{fig:rit_resolution}.}
\end{figure*}
\begin{figure*}[t]
	\center
        \includegraphics[width=0.41\textwidth]{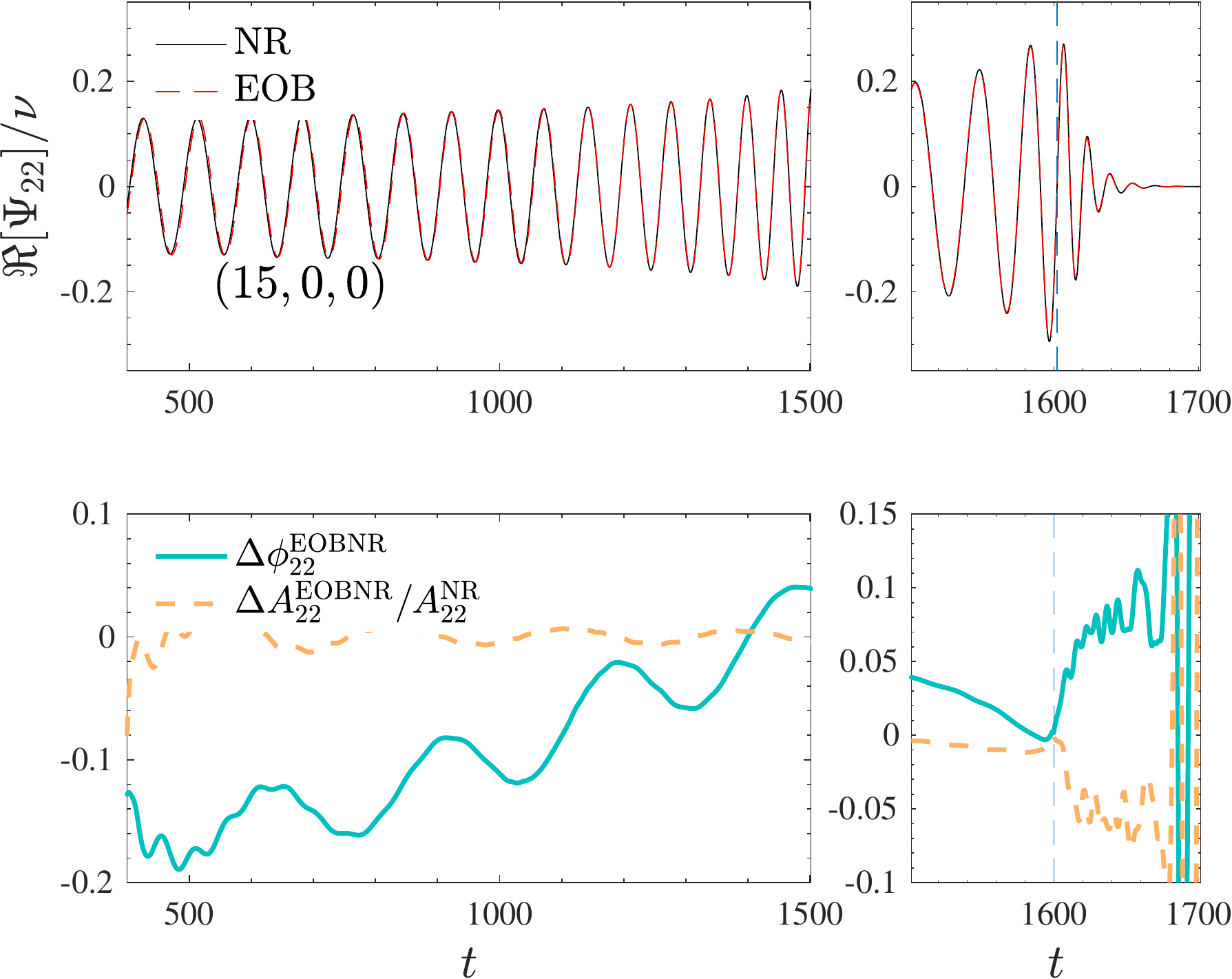}
        \hspace{5mm}
	\includegraphics[width=0.41\textwidth]{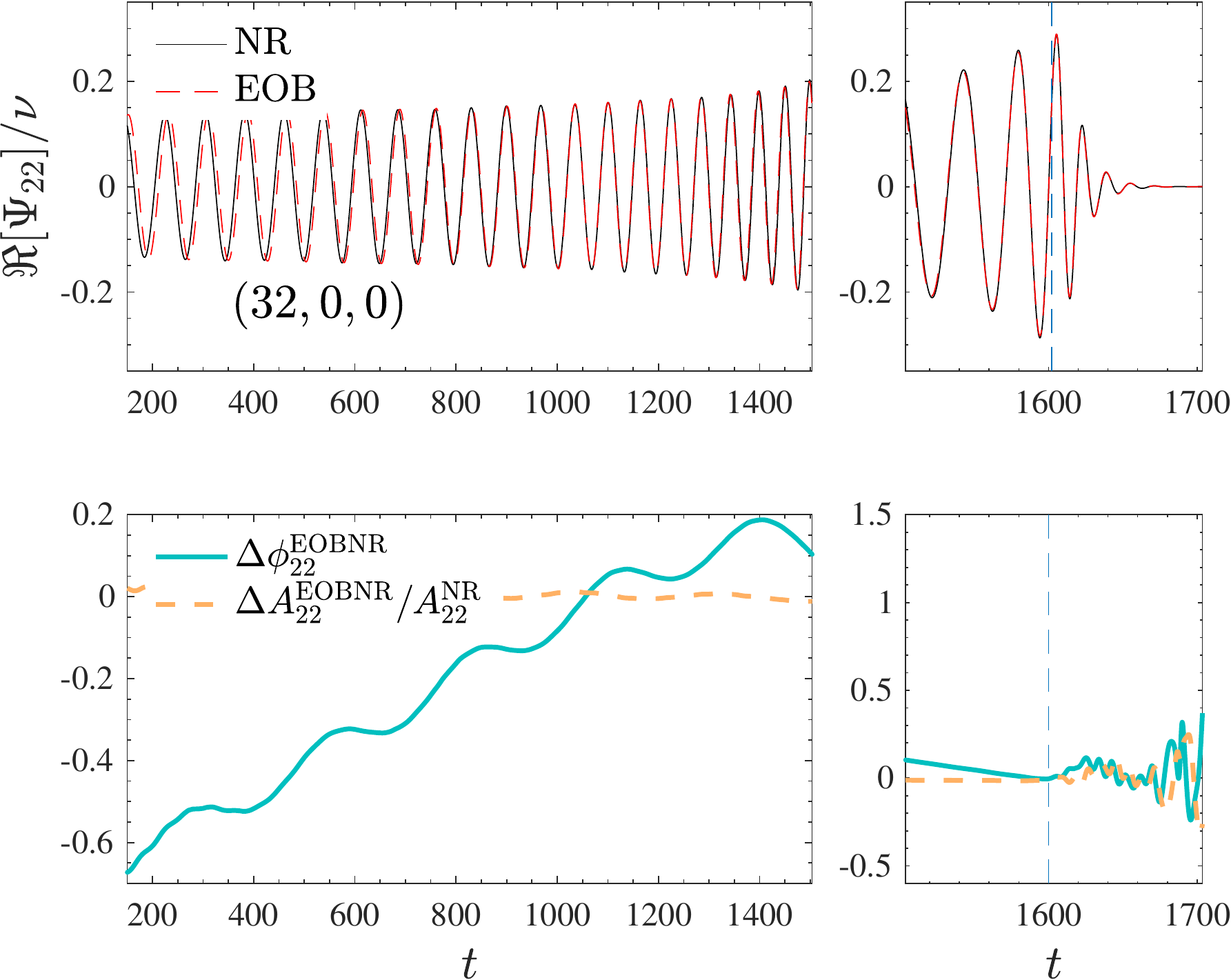}\\
	\vspace{5mm}
	\includegraphics[width=0.41\textwidth]{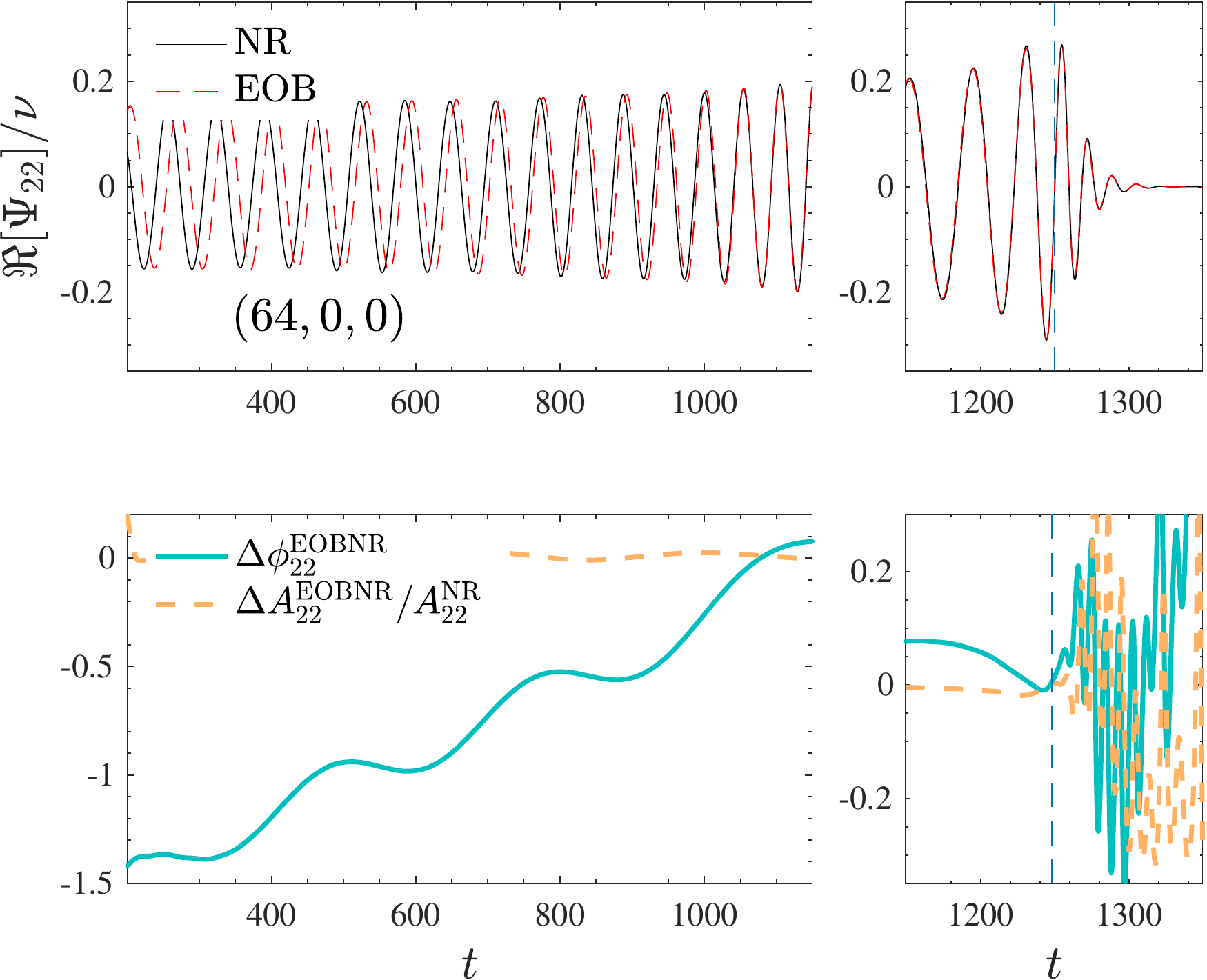}
	\hspace{5mm}
	\includegraphics[width=0.41\textwidth]{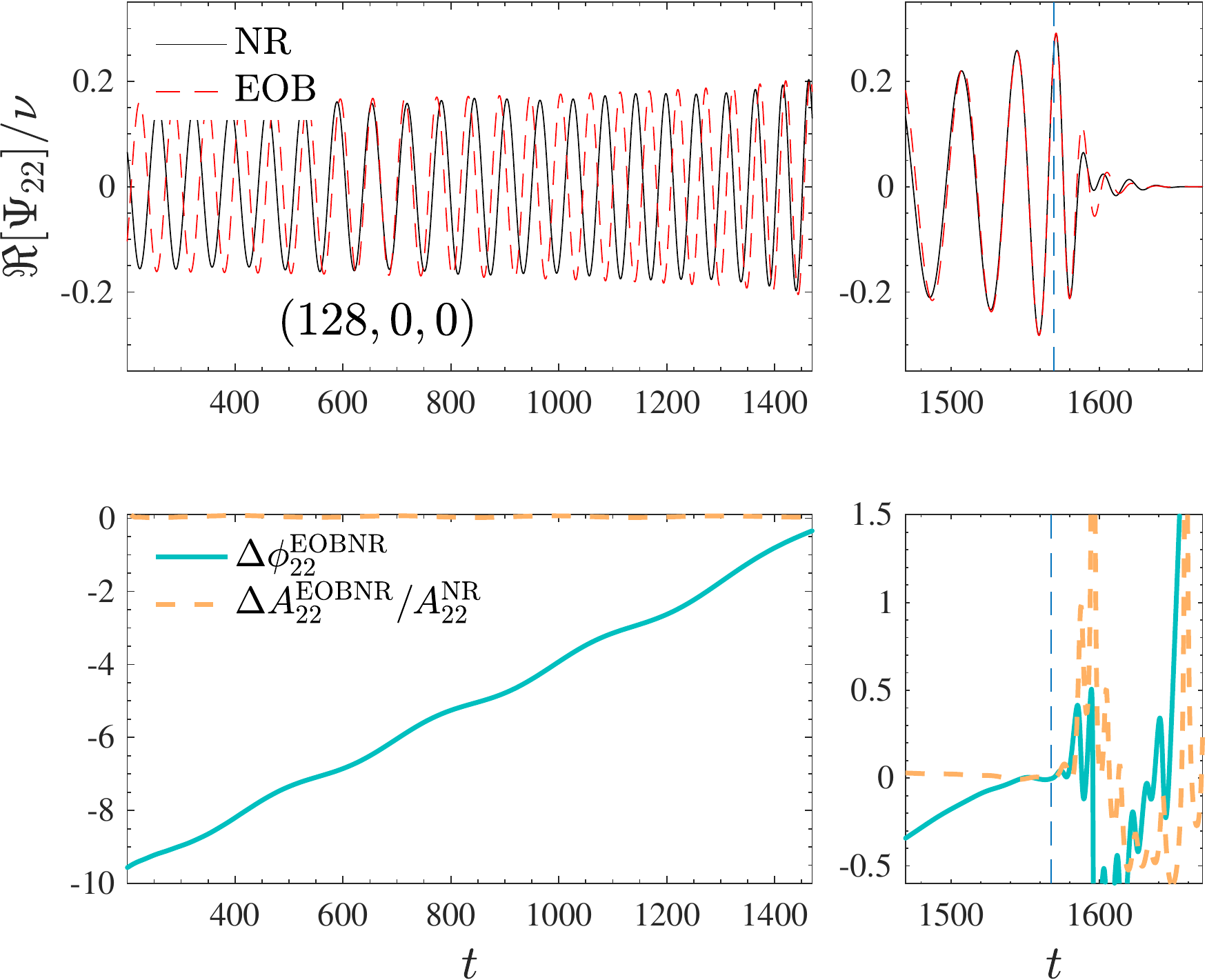}
\caption{\label{fig:eobrit_l2_phasing} EOB/NR-RIT comparison for various mass ratios. For $q=\{15,32\}$ we use resolution-extrapolated
waveforms, while for $q=\{64,128\}$ we adopt the highest resolution available. The waves are aligned in the late inspiral, on
the frequency interval $[\omega_L,\omega_R]=[0.2,0.3]$. For each mass ratio, the upper panel compares the real part of the waveforms,
while the bottom panel shows the phase difference $\Delta\phi^{\rm EOBNR}_{22}\equiv \phi^{\rm EOB}_{22}-\phi^{\rm NR}_{22}$ and
the relative amplitude difference. Note the high EOB/NR consistency during the merger and ringdown phase. 
For $q=128$ the differences in the ringdown are due to NR inaccuracies.}
\end{figure*}
In Figure~\ref{fig:eobrit_l2_phasing} we display the EOB vs. NR-RIT waveform comparisons for a sequence of increasing 
mass ratios, $q=\{15,32,64,128\}$. These naturally show a progressive dephasing and error with increasing $q$.

\subsection{Data at $\mathscr{I}^+$ and memory effect}
\label{sec:spectre}

Additional insight may be given by considering a different type of numerical waveforms provided in
a separate section of the SXS catalog, the Ext-CCE catalog. This section contains asymptotic 
waveforms whose evolution has been run with with SpEC~\cite{SpEC}, and that have been computed in
two ways:
(i) using the Cauchy-characteristic evolution\footnote{%
  As explained in Ref.~\cite{Moxon:2020gha}, 
  Cauchy-characteristic extraction only refers to the transformation from Cauchy coordinates to the set of quantities that 
  are involved in the characteristic evolution. We adopt here their convention and denote as Cauchy-characteristic
  evolution the whole process of Cauchy-characteristic extraction and characteristic evolution.
}
(CCE) scheme implemented in 
SpECTRE~\cite{Moxon:2020gha} and (ii) using the extrapolation procedure
implemented in the python package \texttt{scri}~\cite{Boyle:2013nka,Boyle:2014ioa, Boyle:2015nqa}. 
The latter waveforms are crucially augmented by the nonoscillatory memory contribution as described in 
Mitman et al.~\cite{Mitman:2020bjf}, using a technique that exploits Bondi-Metzner-Sachs (BMS) balance laws. 
This calculation relies on the extraction from the numerical spacetime of the full set of Weyl scalars, 
as discussed in Ref.~\cite{Iozzo:2020jcu}. By contrast, CCE proceeds by exploiting Cauchy data yielded by NR simulations
as a boundary on a timelike worldtube at finite radius, combining it with an exterior evolution on null hypersurfaces
reaching $\mathscr{I}^+$. Though these templates should in principle be more accurate, due to difficulties in choosing
initial data the waveforms of the kind (i) exhibit spurious oscillations, and are hence 
unsuited for our purposes. The latest implementation of the SpECTRE CCE module~\cite{Moxon:2021gbv}
is able to extract waveforms either from finished simulations or from a Generalized Harmonic
simulation simultaneously running in SpECTRE, but this kind of data is currently not readily available.

Interestingly, Ref.~\cite{Mitman:2020bjf} proved the consistency between CCE waveforms and extrapolated waveforms 
improved by the addition of the nonoscillatory memory contribution. The same work also pointed out that 
the memory calculation turns out to be incorrect by $50\%$ for some unknown reason, starting from the 
$\ell=3$, $m=0$ contribution (see especially Sec.~IIIB.2 therein).
Despite these drawbacks and open issues, it is meaningful to perform the same analysis of the $\doublehat{A}_\lm$
quantities using \texttt{scri}-extrapolated waveforms (with the nonoscillatory memory) taken from the Ext-CCE catalog.
In this case, the recommended extrapolation order for the strain $h$ data is $N = 5$.
We focus on $q=4$ nonspinning data: Figure~\ref{fig:A_mrg_spectre}, shows the triple comparison 
between: (i) standard extrapolated $q=4$ SXS data; (ii) the \texttt{scri}-extrapolated data plus the addition of memory and (iii) RIT data.
On the one hand, the figure highlights the consistency between SXS and RIT data. On the other hand, 
the most interesting outcome of the analysis is the {\it much improved consistency} between the test-mass
and $q=4$ scaled amplitudes $\doublehat{A}_{\lm}$ for $\ell=2$, most remarkably for the $\ell=2$, $m=0$ mode.

\section{EOB/NR time-domain phasing comparison}
\label{sec:phasing}

In this section we study the EOB/NR waveform consistency, comparing
higher multipoles and mass ratios to reach to an improvement of some
EOB fitting parameters of particular relevance for large mass ratios.

\subsection{SXS/RIT/EOB consistency for $q=7$}
\label{sec:q=7}
We start our analysis considering a BBH configuration with $q=7$, a mass-ratio regime where 
both SXS and RIT data are well under control. A similar consideration applies to the EOB waveform.
It is thus instructive to drive a triple comparison SXS/EOB and RIT/EOB so to better learn the differences
between the two NR simulations, using the EOB waveform as a reference waveform.
For SXS, we use the SXS:BBH:0298 configuration, taken at the highest available resolution and with 
$N=3$ extrapolation order, since we want to have good control also of the inspiral. 
For what concerns RIT, we  are using resolution extrapolated waveforms so to similarly
minimize the phase uncertainty during the inspiral.
The comparison is shown in Fig.~\ref{fig:sxs_vs_rit}. The waveforms are aligned just before merger time,
using our usual alignment procedure~\cite{Damour:2007yf} that minimizes the EOB/NR phase difference 
in the frequency interval $[\omega_L,\omega_R]=[0.2,0.3]$. The top row of the figure shows the real part of 
the $\ell=m=2$ mode, followed (in the second row) by the phase difference and the relative amplitude difference.
We recall that the RIT waveform is extrapolated in resolution: the EOB/NR phase difference accumulated in
this case is compatible, though larger, than the SXS one, but consistent with the NR uncertainty estimated
in the previous section. The picture also highlights the consistency between ringdowns, although the RIT
one globally looks more accurate, with a slightly smaller phase difference. This might be traced back
to the fact that $(N=2)$-extrapolated  SXS data (more accurate during merger and ringdown) were used 
to construct the ringdown model and not the $N=3$ ones that we are using here. 
Still, the right panel of Fig.~\ref{fig:sxs_vs_rit} proves the reliability and robustness of the NR-fitting procedure
behind the construction of the EOB ringdown model.

The third and fourth rows of Fig.~\ref{fig:sxs_vs_rit} complement the above information showing amplitude
and frequencies for both the $\ell=m=2$ and $\ell=2$, $m=1$ modes. Each panel of the figure incorporates 
several curves: (i) the NR one (black online); (ii) the \teob{} one (red online); (iii) the EOB orbital frequency 
(grey online); (iv) the purely analytical EOB waveform, without NR-tuned next-to-quasi-circular
(NQC) corrections nor NR-informed ringdown (orange online); (v) the curve improved by NQC corrections (light-blue online).
The figure confirms that RIT data are generally closer to the EOB waveform for both modes as well
as it highlights the excellent EOB/NR consistency already achievable with the purely analytical waveform.
An important takeaway message of Fig.~\ref{fig:sxs_vs_rit} is that the presence of a linear-in-time
EOB/NR phase difference for RIT data during the inspiral does not harm the quality of the merger
and ringdown description. This observation will turn out to be useful in the next section, where we will
similarly be analyzing RIT data with larger mass ratios.

This general good agreement of SXS and RIT NR waveforms supplements those
observed for the sources of GW150914~\cite{Lovelace:2016uwp} and GW170104~\cite{Healy:2017xwx} 
for the more comparable mass ratios and up to $\ell=4$ modes.

\subsection{RIT/EOB comparison for large mass ratios}
\label{sec:large_q}
Let us focus now on the $\ell=m=2$ EOB/NR phasing comparisons for $q=\{15,32,64,128\}$.
Likewise the $q=7$ case, waveforms are aligned just before merger.  For $q=\{15,32\}$ we provide 
comparisons with the resolution-extrapolated waveform, while for $q=\{64,128\}$ we use the 
highest resolution available. The plot show a remarkable EOB/NR agreement during
merger and ringdown, despite not having used any of this data to inform \teob{}.
The secular EOB/NR dephasing accumulated during the inspiral is related to the finite resolution
of the simulation and it is of no concern at the moment. Note in particular that for $q=15$, the
phase difference accumulated towards early frequency is $\Delta\phi^{\rm EOBNR}_{22}\sim -0.2$~rad,
that is of the order of the estimated NR uncertainty. The accumulated $\Delta\phi^{\rm EOBNR}_{22}$ 
is at most of the order of $\sim 1$~rad  up to $q=64$. Seen the coherence between $q=\{15,32,64\}$ we
think that this value is consistent with a (conservative) error estimate of the NR uncertainty (especially
considering that $q=64$ data are not extrapolated in resolution) and thus we can claim that NR data,
in a sense, are loosely testing also the radiation-reaction dominated epoch of the waveform up to
$q=64$.  By contrast, this statement is certainly not correct for $q=128$, that is a much more demanding
simulation. Higher resolution will be probably needed here to mutually test the two approaches in this
regime. For the moment, we think we can claim that \teob{} is here giving the most accurate
(approximate) representation we have for an inspiral waveform of a $q=128$ BBH.

\subsection{Higher multipolar waveform modes}
\label{sec:hm}
\begin{figure}[t]
	\center
        \includegraphics[width=0.45\textwidth]{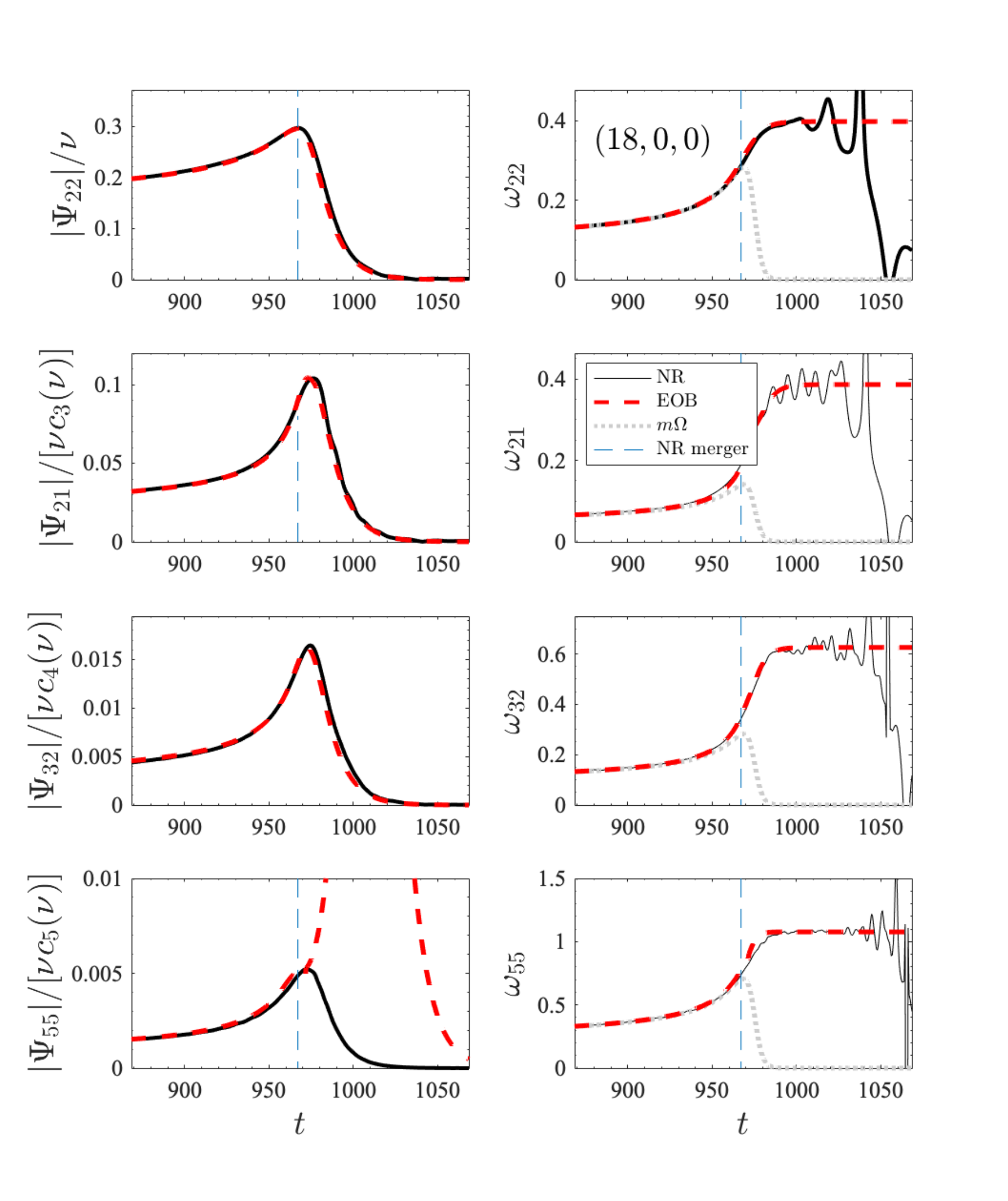}
\caption{\label{fig:q18}EOB/NR comparison for $q=18$ using the NR data of Ref.~\cite{Husa:2015iqa}.
The unphysical behavior of the $\ell=m=5$ mode after merger is evident.}
\end{figure}

Let us finally complete our analysis considering higher modes. The \teob{} modes completed through
merger and ringdown are $(2,1)$, $(3,3)$, $(3,2)$, $(4,4)$, $(4,3)$ and $(5,5)$. When tested all over
the (nonspinning) parameter space, all modes are generated robustly, without evident 
pathological features, except for the $(5,5)$ one. This mode displays unphysical behavior already 
for $q\gtrsim 15$. This was  already noted by one of us during the first development of the model,  
in Ref.~\cite{Nagar:2019wds} while driving comparisons with a $q=18$ NR dataset obtained using
the BAM code from Ref.~\cite{Husa:2015iqa}, although not explicitly reported. 
Figure~\ref{fig:q18}  is an EOB/NR amplitude and frequency comparison using 
the $q=18$ BAM data  of Ref.~\cite{Husa:2015iqa}. This complements the $\ell=m=2$ mode 
comparison shown in Fig.~12 of Ref.~\cite{Nagar:2019wds}. The picture highlights  the 
incorrect behavior of the $\ell=m=5$ mode amplitude after merger. Similarly, the analytical 
frequency does not match the NR one. Building an improved analytical description of
the $\ell=m=5$ mode will be the subject of the next section.

\begin{table}[t]
 \caption{\label{tab:DTlm} Time delay  $\Delta t_{\lm}$ between the $\ell=m=2$ NR waveform peak and the corresponding 
 peaks for $\ell=m$ multipoles up to $\ell=6$. The second column indicates the  identification number of each simulation 
 from the SXS and RIT waveform catalog. Exception to this are the $q=18$ dataset, obtained in Ref.~\cite{Husa:2015iqa} 
 using the BAM code, and the test-mass limit waveform obtained using {\tt Teukode}~\cite{Harms:2014dqa,Harms:2013ib}.}
   \begin{center}
     \begin{ruledtabular}
\begin{tabular}{c l c c c c c} 
$\#$ & ID & $q$ & $\nu$ & $\Delta t_{44}$ &  $\Delta t_{55}$  & $\Delta t_{66}$ \\
  \hline
  \hline
  1 & SXS:BBH:1153 & 1      & 0.25  & 3.6587  & $\dots$ & 6.65044\\
  2  & SXS:BBH:0198         & 1.203 &       0.2479  & 3.4943  & $\dots$ & $\dots$ \\
  3    &  SXS:BBH:1354  & 1.832  & 0.2284 & $\dots$  & 4.8445 & $\dots$ \\
  4   &    SXS:BBH:1166                        & 2        &  $0.\bar{2}$ & $\dots$ & 4.4172 & $\dots$\\
  5   & SXS:BBH:0191  & 2.507    &  0.2038  & 2.1388 & \dots & $\dots$ \\
  6 &  SXS:BBH:1178 & 3          &  0.139 & 1.601 & 4.196 & $\dots$ \\
  7 & SXS:BBH:0197  & 5.522     &  0.0988 & 3.6521& 4.6133&4.520 \\ 
  8 & SXS:BBH:0298  & 7  & 0.1094& 3.7126 & 4.6045 &  5.2422 \\
  9 & RIT:BBH:0416     & 7  & 0.1094& 4.2687& 4.6794 & 4.6301 \\
  10 & SXS:BBH:0301          & 9  & 0.09& 4.2998& 5.2108 & 5.7425\\
  11 & SXS:BBH:1107          & 10  & 0.0826& 4.3957 & 5.3862 & 6.087 \\
  12    & RIT:BBH:0373                             & 15 & 0.0586 &4.34 & 4.7081 & 4.658\\
  13   & BAM~\cite{Husa:2015iqa}    & 18 & 0.0499& 4.4054 & 5.1464 & 5.8734\\
  14     & RIT:BBH:0792    & 32 &0.0294 &2.8970 &2.9929 &2.0056\\
  15    & RIT:BBH:0812    & 64 &0.0151 &2.946 & 2.7026 & 1.996\\
  16     & RIT:BBH:0935   & 128 &0.0077 &3.524 & 5.0108 &4.4429\\
  17 &  {\tt Teukode} & $\infty$ & 0 & 5.2828 & 6.5618 & 7.7 \\ 
  \end{tabular}
 \end{ruledtabular}
 \end{center}
 \end{table}
\begin{figure*}[t]
	\center
        \includegraphics[width=0.45\textwidth]{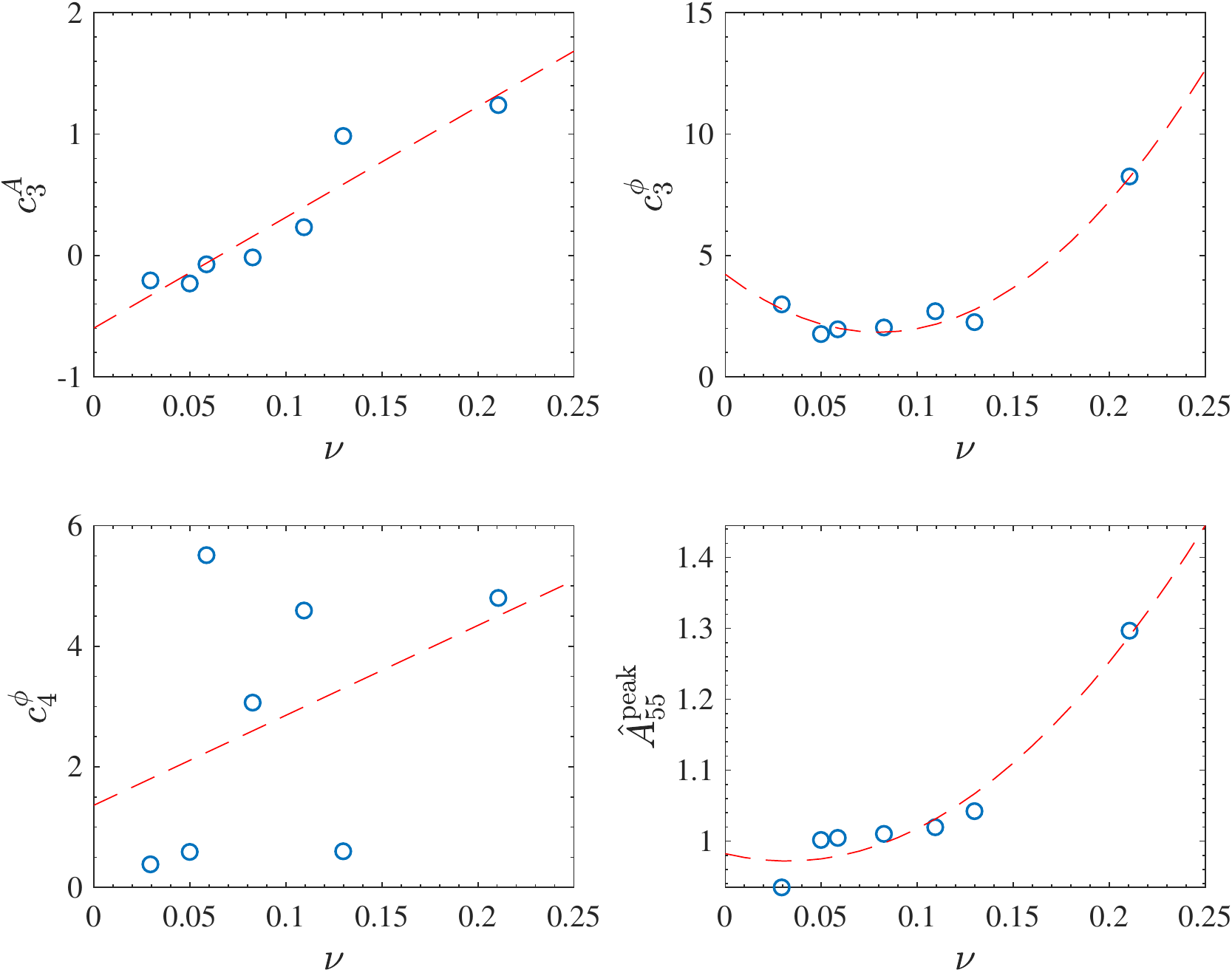}
        \hspace{5mm}
	\includegraphics[width=0.45\textwidth]{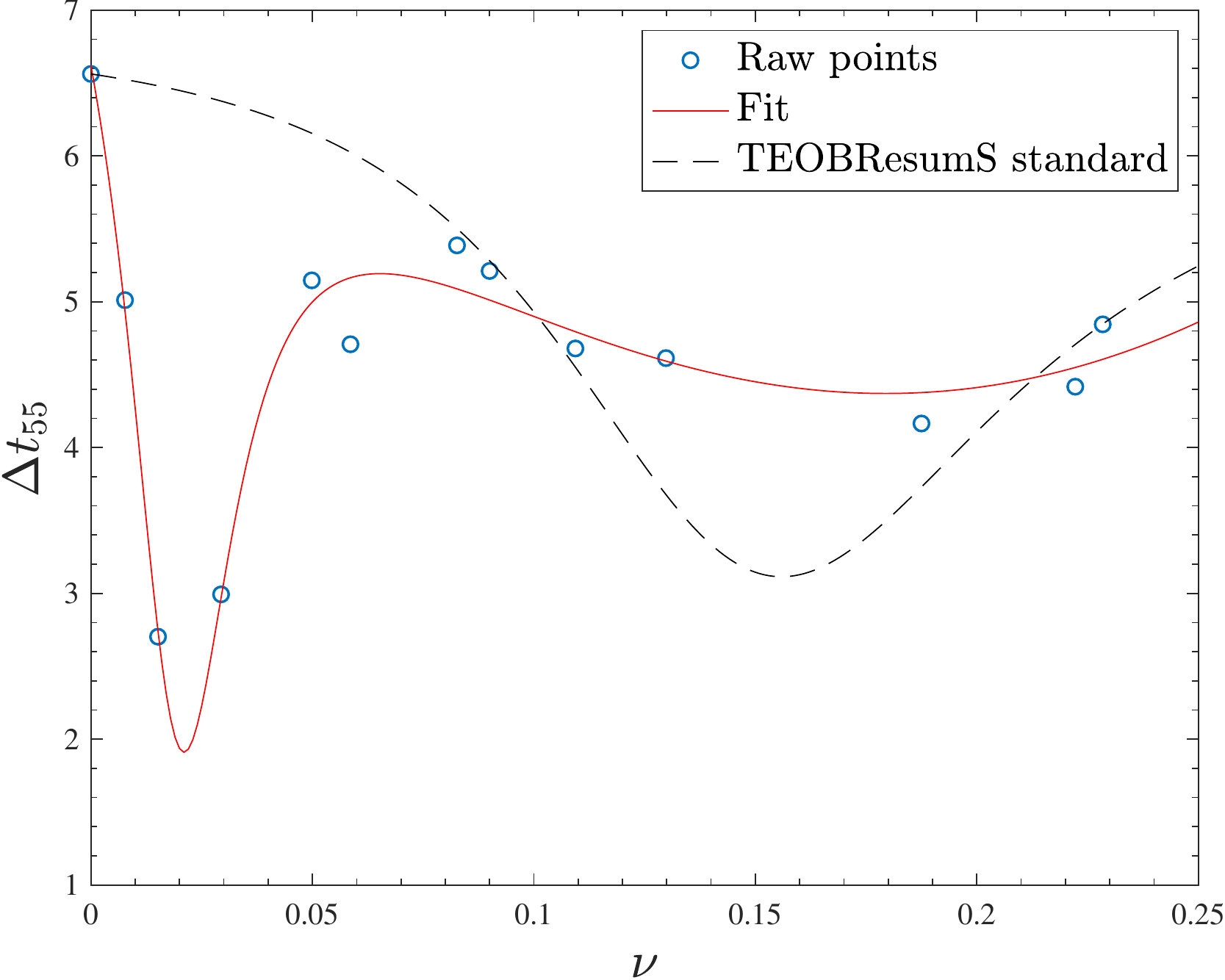}	
\caption{\label{fig:fits}NR points and interpolating fits for the quantities entering the postpeak description of the $(5,5)$ mode.
Note the special behavior of the $\Delta t_{55}$ that is necessary to correctly capture the values for $q=32$ and $q=64$. 
The previously implemented fit (incorrectly informed by other SXS datasets) is superposed for completeness.}
\end{figure*}
\begin{figure}[t]
	\center
        \includegraphics[width=0.23\textwidth]{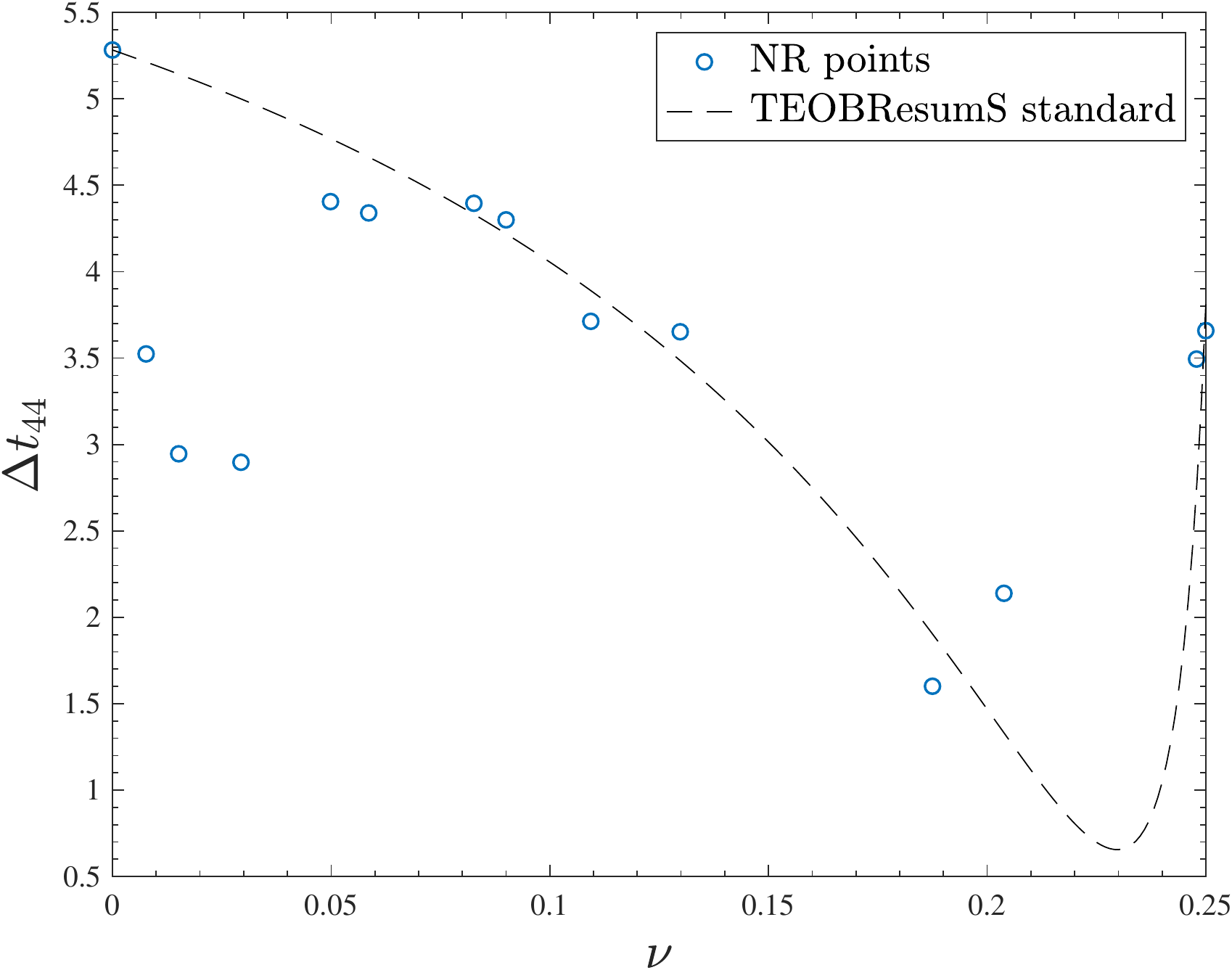}
        \hspace{2mm}
	\includegraphics[width=0.23\textwidth]{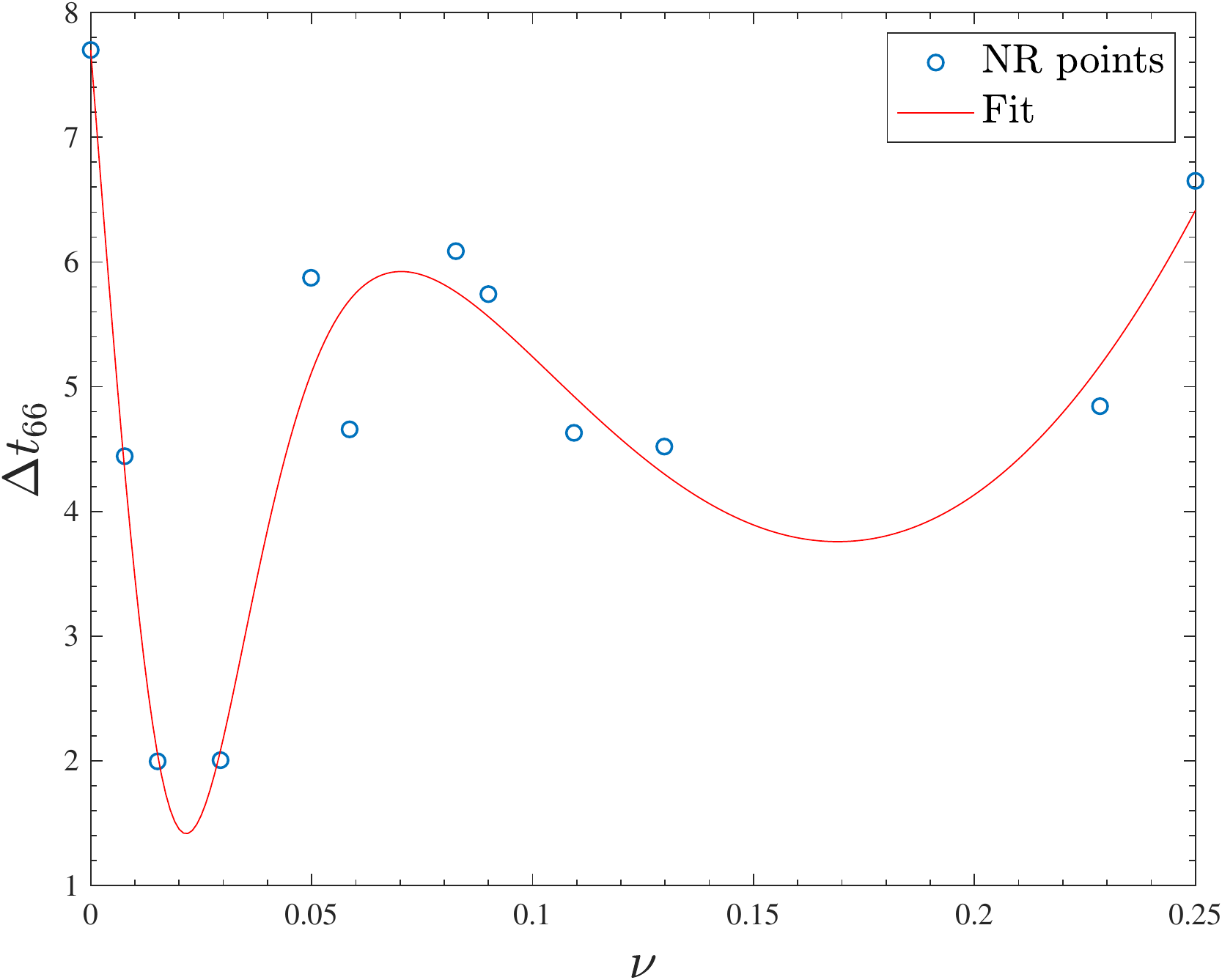}	
\caption{\label{fig:fits_44_66}Behavior of the $\Delta t_{44}$ and $\Delta t_{66}$ points of Table~\ref{tab:DTlm}. For
the $\Delta t_{44}$ we also superpose the current fit implemented in \teob{}. The differences with the NR data 
for small values of $\nu$ do not seem to impact the EOB/NR waveform agreement (see Fig.~\ref{fig:eobrit_HM}).
The behavior of $\Delta t_{66}$ is qualitatively analogous to the $\Delta t_{55}$ one.}
\end{figure}
\begin{figure*}[t]
	\center
	\includegraphics[width=0.46\textwidth]{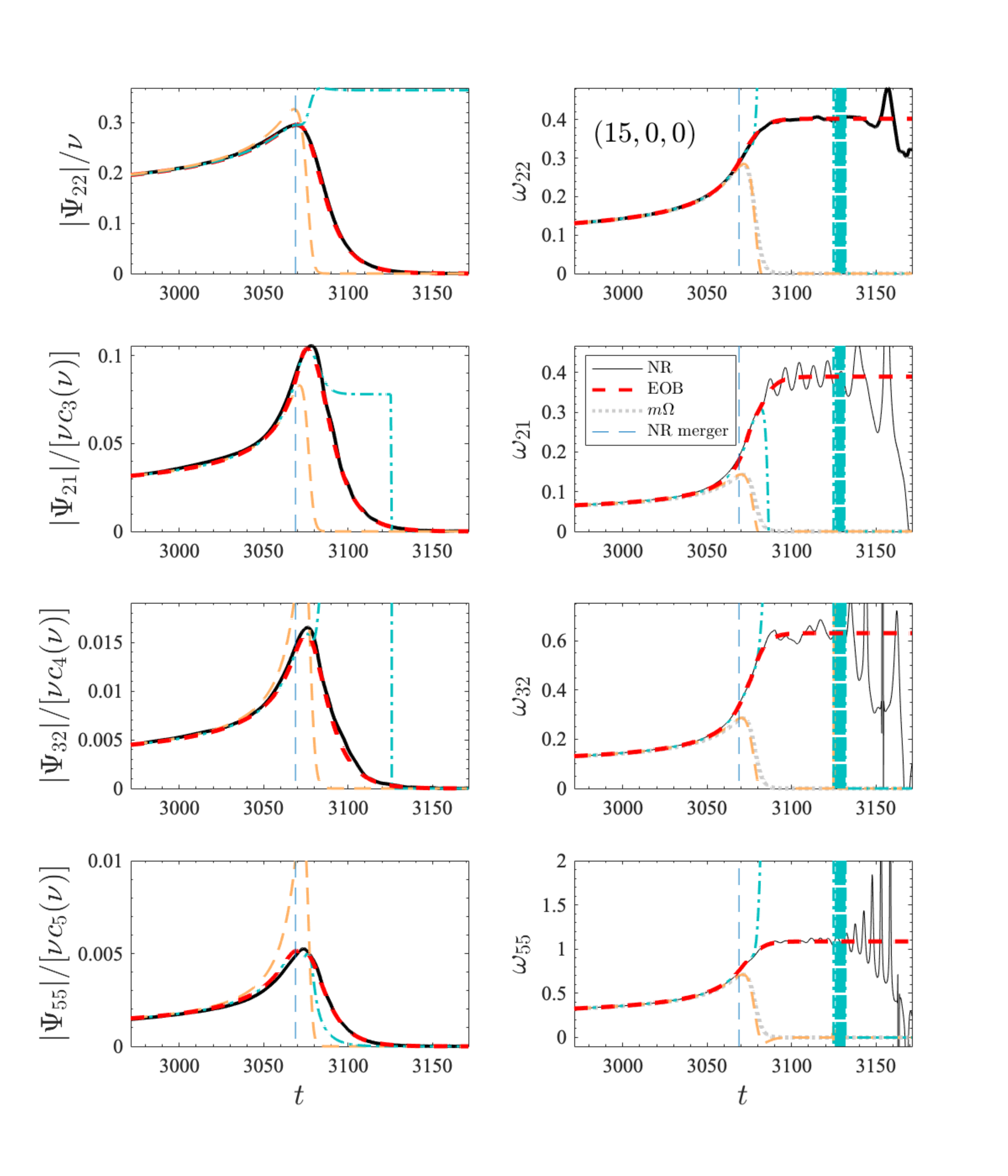}
	\includegraphics[width=0.46\textwidth]{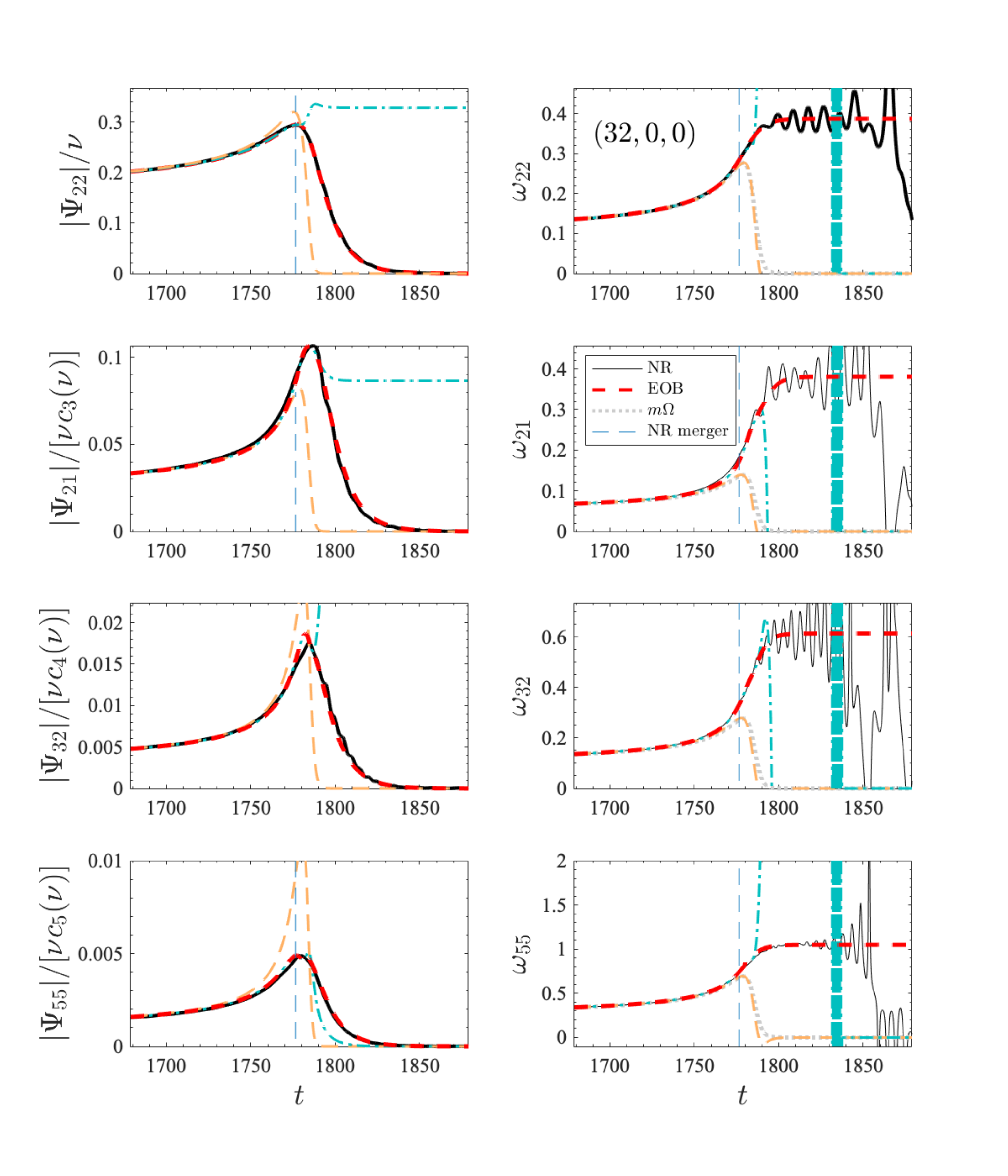}\\
	\vspace{-8mm}
	\includegraphics[width=0.46\textwidth]{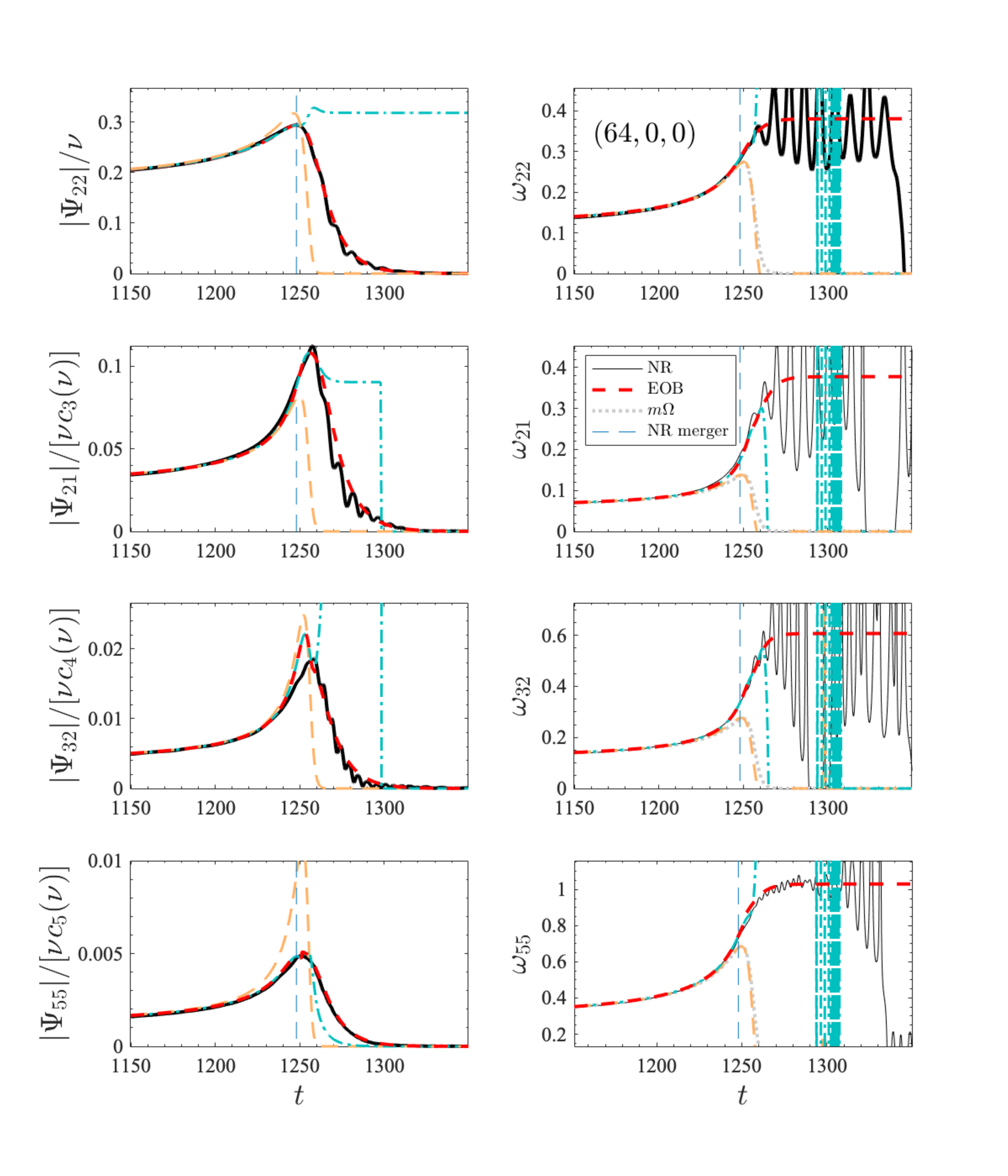}
	\includegraphics[width=0.46\textwidth]{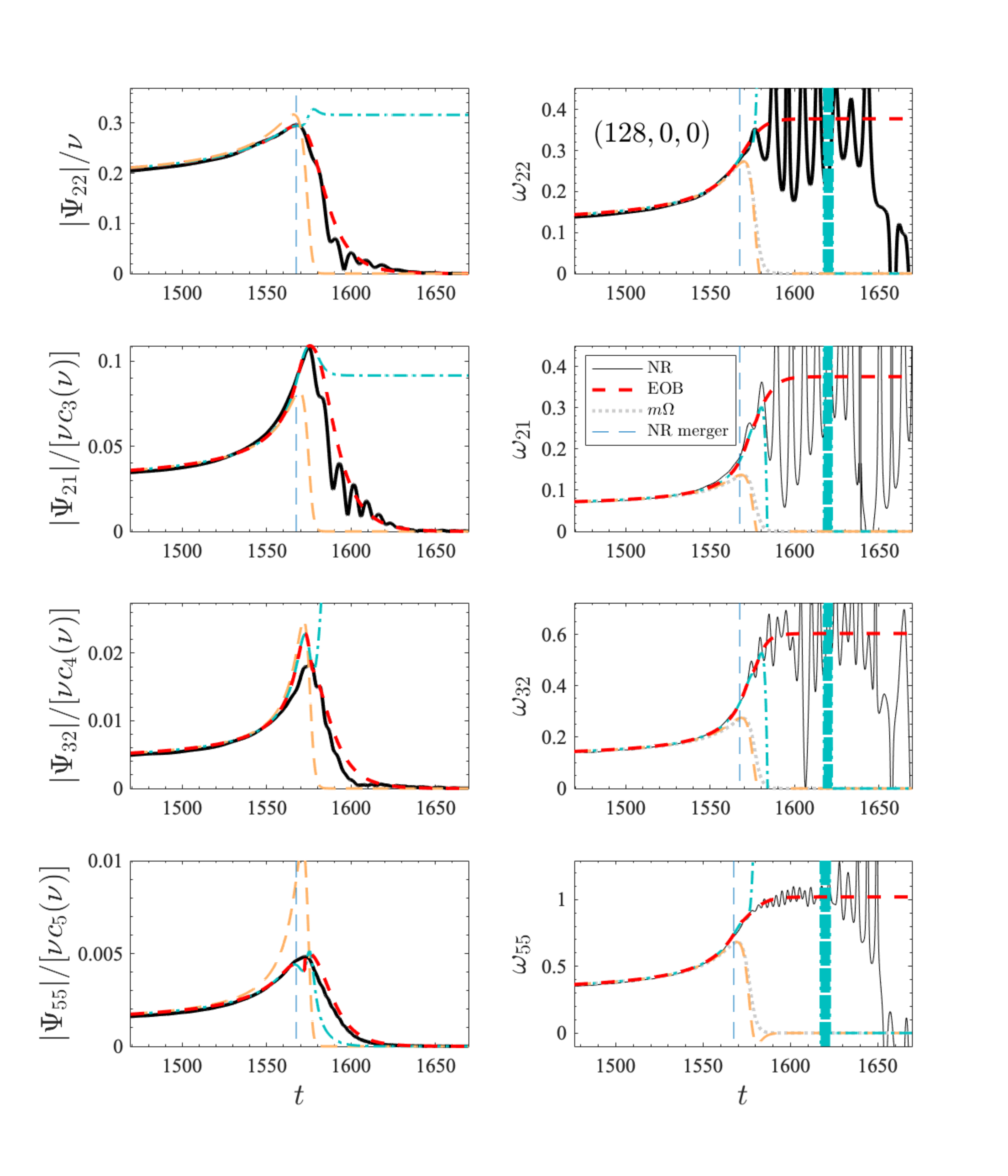}
\caption{\label{fig:eobrit_HM} EOB/NR comparison of amplitude and frequency including higher modes. The $\ell=m=5$
mode uses the new ringdown fit described in Sec.~\ref{sec:l5m5}. We report the analytical waveform (orange online),
the NQC-completed waveform (light-blue online) and the one completed through merger and ringdown (red online).}
\end{figure*}

\begin{figure*}[t]
	\center
	\includegraphics[width=0.46\textwidth]{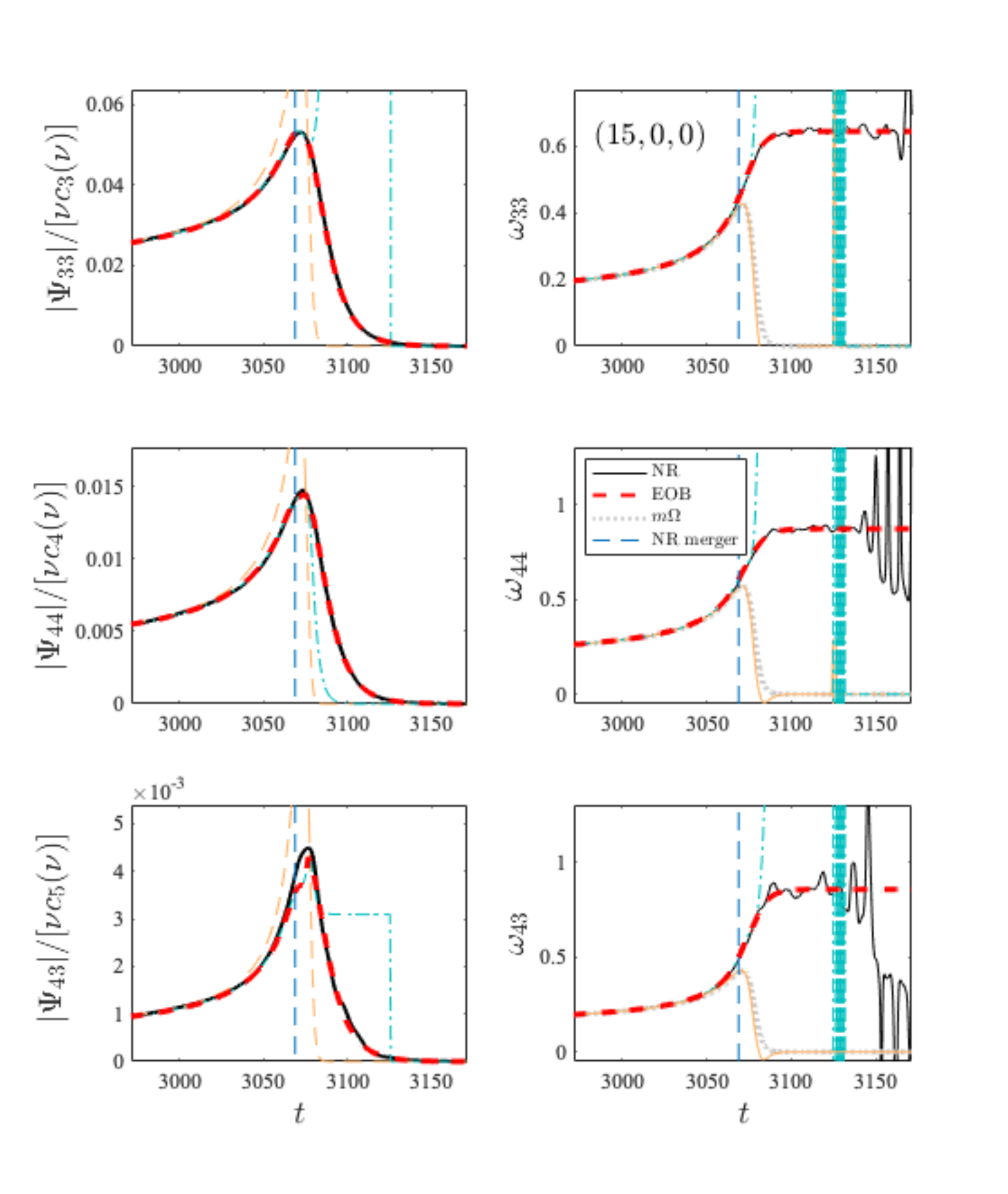}
	\includegraphics[width=0.46\textwidth]{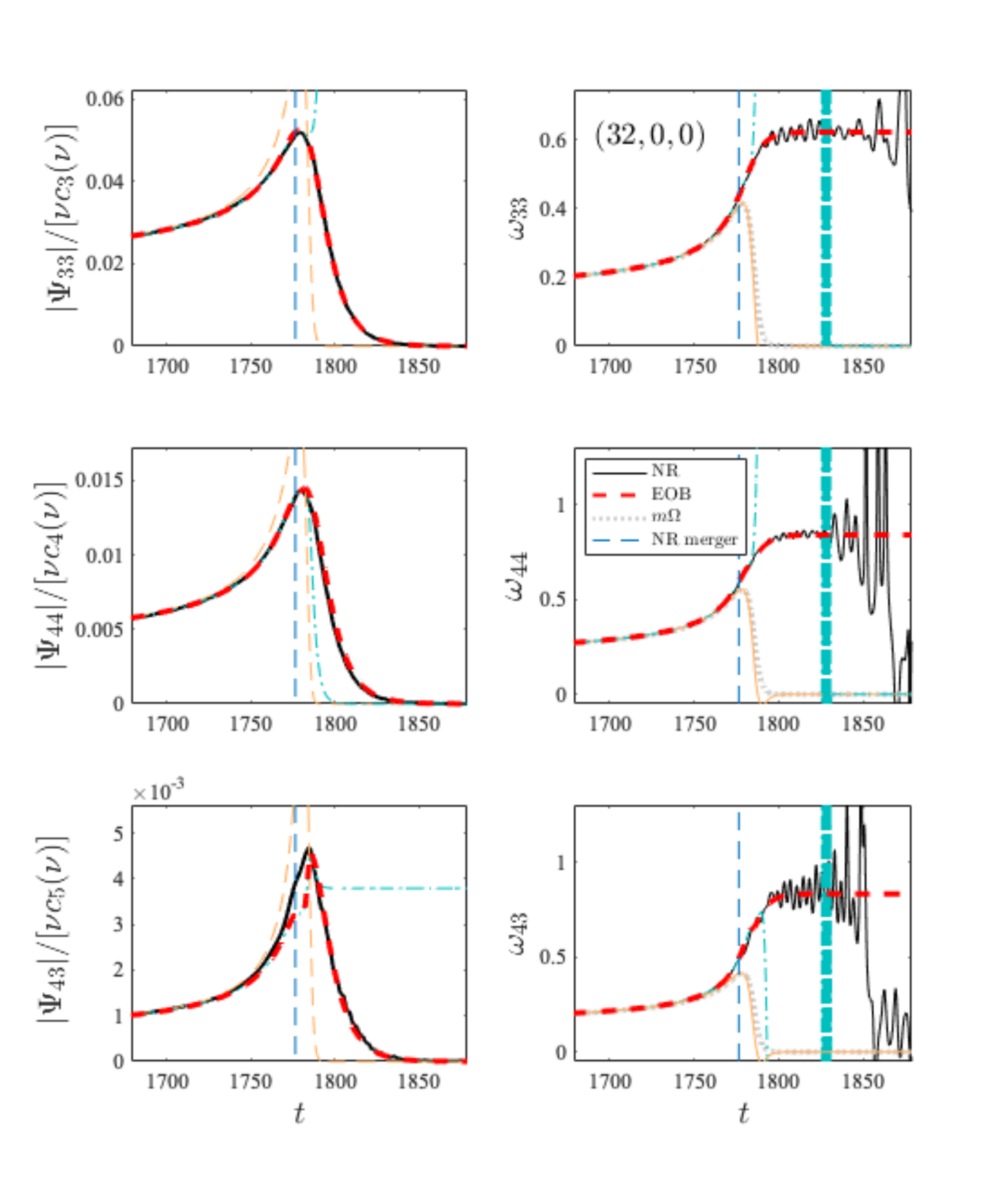}\\
	\vspace{-8mm}
	\includegraphics[width=0.46\textwidth]{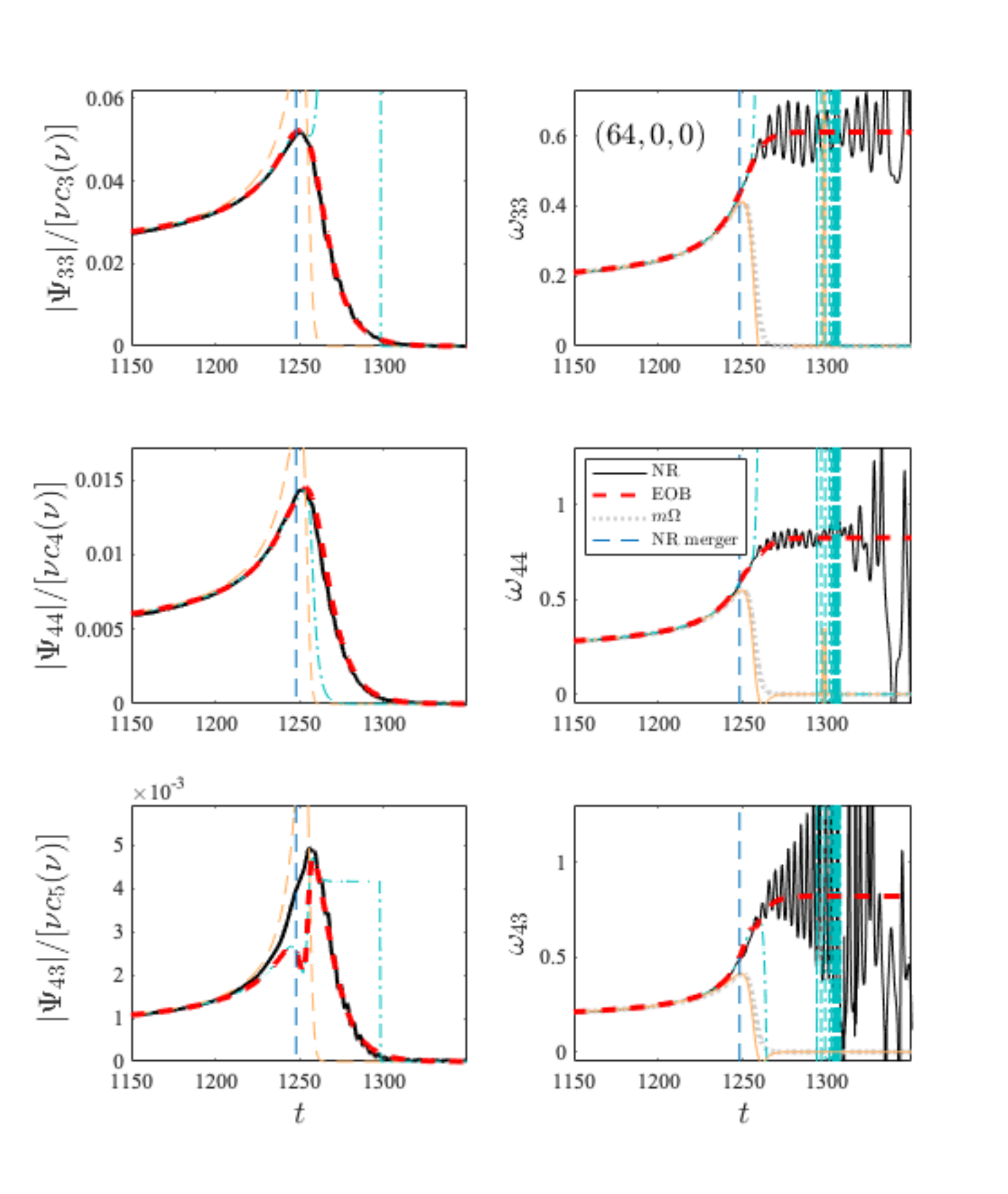}
	\includegraphics[width=0.46\textwidth]{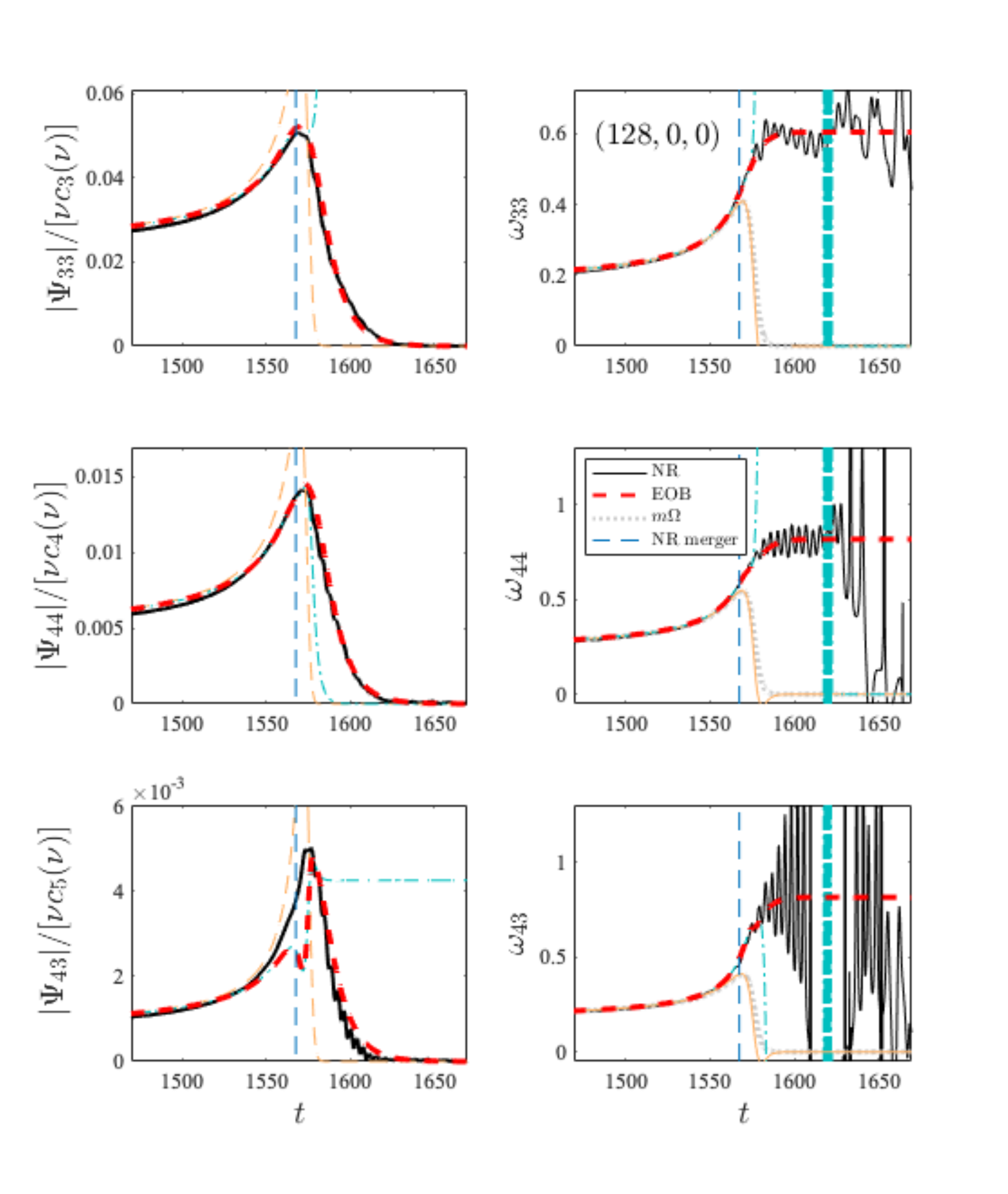}
\caption{\label{fig:eobrit_HM_43} EOB/NR comparison of amplitude and frequency including the higher modes not considered in Fig.~\ref{fig:eobrit_HM}.
Note that the $(4,3)$ amplitude becomes increasingly inaccurate before merger (due to the imperfect action of the next-to-quasi-circular factor) 
as $q$ increases.We report the analytical waveform (orange online), the NQC-completed waveform (light-blue online) and the 
one completed through merger and ringdown (red online).}
\end{figure*}

\subsubsection{Improved analytical description of the $\ell=m=5$ merger and ringdown waveform}
\label{sec:l5m5}

The ringdown (or better saying, the {\it postpeak}) description of each multipole within \teob{} is based on the NR-informed 
fitting procedure introduced in Ref.~\cite{Damour:2014yha}. This approach, originally discussed for the $\ell=m=2$ mode,
was extended to higher modes and gives one of the essential building blocks of \teob~\cite{Nagar:2019wds,Nagar:2020pcj}.
The method also yields a stand-alone time-domain waveform model that can be used in targeted ringdown 
analyses~\cite{DelPozzo:2016kmd,Carullo:2018gah}, and improvement in the modeling of the 
amplitude already exists~\cite{Albanesi:2021rby}. 
Here we build upon Ref.~\cite{Nagar:2019wds} and improve the (nonspinning) fits for the $\ell=m=5$
postpeak waveform presented there. To do so, we (i) use a new sample of carefully chosen SXS datasets, 
with $N=2$ extrapolation order and mass ratio $1<q\leq 10$; (ii) complement this data with a $q=18$ 
BAM waveform already used in previous work~\cite{Nagar:2019wds} and all 
the $q=\{15,32,64,128\}$ datasets discussed above.
This is essential to correctly connect the comparable-mass regime with the extreme-mass-ratio limit.
We present here new fits for the amplitude peak $\hat{A}_{55}$, for the three parameters $(c^3_A,c^3_\phi,c^4_\phi)$ 
entering the  postpeak description (see Ref.~\cite{Damour:2014yha} for details),  and for $\Delta t_{55}$, 
the time lag between the peaks of the $(2,2)$ and $(5,5)$ modes. This quantity is especially 
important because it is the one that assures that the postpeak waveform is attached to the inspiral waveform 
at the correct point.

Following Ref.~\cite{Nagar:2019wds}, both $\hat{A}_{55}$ and $\Delta t_{55}$ are fitted after factorization
of their values in the test-mass limit and, for the amplitude, of the leading-order dependence on $\nu$.
We use different SXS simulations depending on the quantity we have to fit.
In the left panel of Fig.~\ref{fig:fits} we show the raw points for $\hat{A}_{55}$ and $(c^3_A,c^3_\phi,c^4_\phi)$,
extracted from  $q=\{1,1.5,2,3,4,5,6,8,9.89\}$ SXS simulations, with the best-fit functions superposed.
They are given by
\begin{align}
\hat{A}_{55}^{\rm peak} &= 1- 0.97509\nu + 11.20088\nu^2 , \\
c^3_A     &=  -0.59703 + 9.11875\nu ,\\ 
c^3_\phi   &=  4.22624-59.69283\nu+ 373.31260\nu^2 \ ,\\ 
c^4_\phi   &= 1.36397 + 14.911137\nu  .   
\end{align}
The accurate representation of $\Delta t_{55}(\nu)$ is a crucial element to assure that the postpeak 
waveform is attached at the correct place. As a consequence, we were especially careful in selecting
the NR datasets that are listed in Table~\ref{tab:DTlm}. We selected the SXS simulations under the
conditions that the values are {\it stable} with $\nu$, i.e. small variations of $\nu$ yield small variations
in $\Delta t_{55}$. This is not always the case when using data in the SXS catalog and extra care should
be exerted in the dataset choice, since the $(5,5)$ mode seems particularly sensitive to the
appearance of unphysical effects. The points of Table~\ref{tab:DTlm} are shown in the right panel
of Fig.~\ref{fig:fits}. We see that the behavior of $\Delta t_{55}$  for $q=(32,64,128)$ is rather complicated
and it is necessary to have this data in order to correctly enforce the $\nu\to 0$ limit.
The fit for $\Delta t_{55}$ reported in the figure explicitly reads
\be
\Delta t_{55}=n_0\dfrac{1 + n_1\nu +n_2\nu^3 + n_3\nu^3 + n_4 \nu^4}{1 + d_1\nu + d_2\nu^2} \ ,
\ee
where
\begin{align}
n_0 &= 6.6195,\\
n_1 &=-91.2039,\\
n_2 &= 2556.5123,\\
n_3 &= -11325.217,\\
n_4 &= 27767.2164,\\
d_1 &= -66.0362,\\
d_2 &= 1762.4169.
\end{align}
Let us finally briefly comment on the $\Delta t_{44}$ and $\Delta t_{66}$ points listed in Table~\ref{tab:DTlm}.
The $\Delta t_{44}$ points are shown in the left panel of Fig.~\ref{fig:fits_44_66} together with the fit
of Ref.~\cite{Nagar:2020pcj}, that is currently implemented in \teob{} (dashed line on the plot). We see
that, similarly to the $\Delta t_{55}$ case, $\Delta t_{44}$ shows a special behavior for small $\nu$ that
is not captured by the fit. Since we have found that this does not have a relevant influence
on the modelization of the $\ell=m=4$ mode for large mass ratios (see the corresponding plots in 
Fig.~\ref{fig:eobrit_HM} below), we have decided to keep the standard \teob{} fit. The $\Delta t_{66}$ 
points display the same qualitative behavior of the $\Delta t_{55}$ ones, and thus it is necessary to use a 
sufficiently flexible rational function to fit them robustly (solid line in the plot). 
In conclusion, our analysis shows that NR simulations of large mass ratio binaries encode important 
information that needs to be taken into account so that the \teob{} model correctly tends to the test-mass 
limit. Simpler interpolations to the test-mass limit can eventually introduce systematic effects that may 
invalidate robust performance all over the parameter space.

\subsubsection{Global comparison}
Figure~\ref{fig:eobrit_HM} and~\ref{fig:eobrit_HM_43} illustrate the EOB/NR agreement around merger for all modes that are robustly 
completed through merger and ringdown, i.e.: $(2,2)$, $(2,1)$, $(3,3)$, $(3,2)$, $(4,4)$, $(4,3)$ and $(5,5)$.
Note that here, to ease the comparison, we are not using resolution-extrapolated waveform data, but highest resolution data instead. 
The reason for this choice is that the extrapolation process can fictitiously 
magnify the oscillations in the frequency that are present during the ringdown in some modes, e.g. the (2,1) mode.
Analogously to the $q=7$ case mentioned above, for each mass-ratio and mode reported in Fig.~\ref{fig:eobrit_HM} 
we compare four curves: (i) the NR one (black online); (ii) the purely analytical one (orange online); (iii) 
the waveform augmented with NQC corrections (light-blue online) and (iv) the full waveform completed 
with merger and ringdown (red online). We note now the robustness of the $\ell=m=5$ mode, that is modeled 
using the new NR-informed ringdown fits described above.
Note however that some unphysical features appear in the $(4,3)$ mode amplitude as the mass ratio is 
increased. In this case, the feature is coming from the NQC correction to the amplitude, while the behavior 
during ringdown is robust and consistent with the NR waveform for {\it any} mass ratio. The improvement
of the $(4,3)$ mode for large values of the mass ratio will require a new NQC-determination strategy that
will be investigated in future work.

\section{Informing EOB models using NR simulations: hunting for systematics}
\label{sec:EOBNR}
\begin{figure}[t]
	\center
	\includegraphics[width=0.42\textwidth]{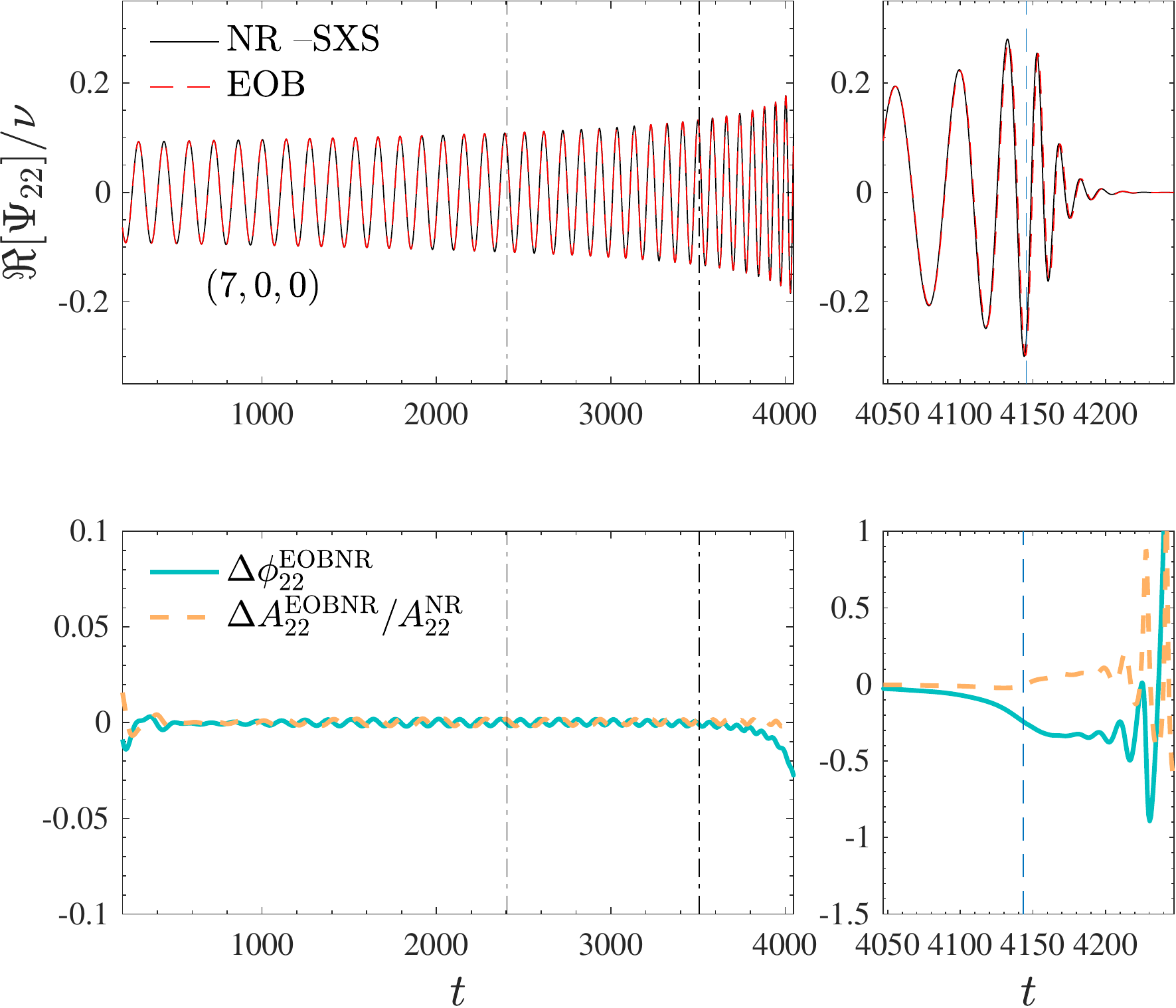}
\caption{\label{fig:q7_sxs}EOB/NR phasing comparison with SXS:BBH:0298 ($q=7$, nonspinning) extrapolated with $N=3$. 
The vertical dash-dotted lines indicate the alignment region. The phase difference at merger point, $\sim -0.25$~rad, is consistent with 
(and notably larger than) the NR phase uncertainty $\delta\phi_{\rm mrg}^{\rm NR}=-0.0775$~rad.}
\end{figure}

\begin{figure}[t]
	\center
	\includegraphics[width=0.42\textwidth]{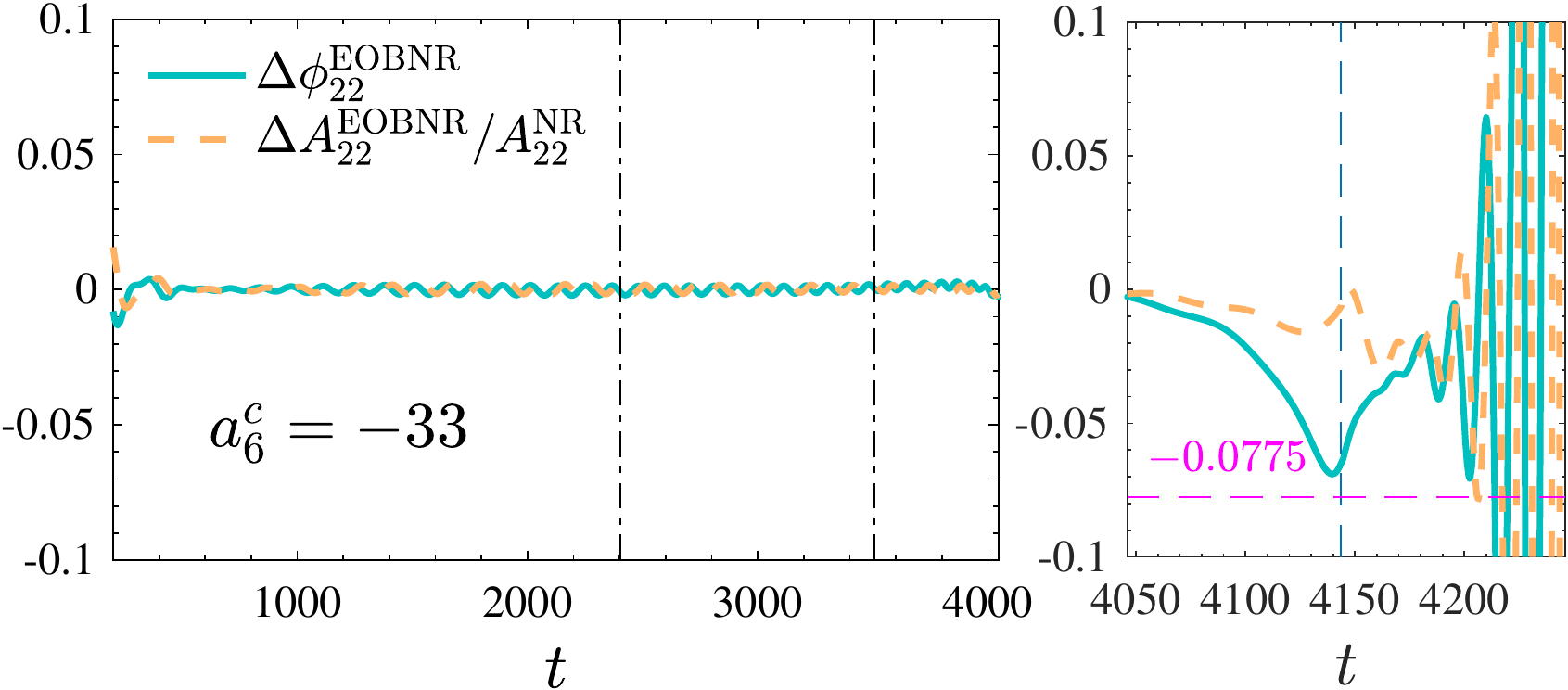}
\caption{\label{fig:q7_sxs_m33}Same EOB/NR phasing comparison with SXS:BBH:0298 of Fig.~\ref{fig:q7_sxs} but using
now $a_6^c=-33$ instead of the value given by Eq.~\eqref{eq:a6c}: the EOB/NR phase difference at merger is now slightly 
smaller than the NR phase uncertainty $\delta\phi_{\rm mrg}^{\rm NR}=-0.0775$~rad. }
\end{figure}

EOB analytical waveform models are informed by NR simulations. The idea of incorporating in the model 
strong-field bits of information extracted from NR was suggested already two decades ago~\cite{Damour:2002qh},
at the dawn of the EOB development. Nowadays, NR-informing EOB models is a crucial step to make them
highly faithful with respect to, error-controlled, NR waveform data~\cite{Damour:2009kr,Nagar:2015xqa,Bohe:2016gbl,Cotesta:2018fcv,
Nagar:2019wds,Nagar:2020pcj,Nagar:2021gss,Nagar:2021xnh,Riemenschneider:2021ppj,Albertini:2021tbt}.
In particular, the spin-aligned \teob{} model incorporates NR information in: (i) the ringdown part, as discussed above; 
(ii) the NQC corrections to the waveform; (iii) an effective 5PN function $a_6^c(\nu)$ entering the orbital interaction 
potential $A(r;\nu)$, i.e. the $\nu$-dependent deformation of the Schwarzschild potential $1-2/r$; 
(iv) an effective next-to-next-to-next-to-leading order (i.e. at 4.5PN accuracy) function $c_3(\nu)$ entering 
the spin-orbit coupling term of the Hamiltonian. Here we are only dealing with nonspinning configurations, so our interest 
is limited to $a_6^c(\nu)$. Reference~\cite{Nagar:2019wds} used several SXS datasets to determine $a_6^c(\nu)$ as
\be
\label{eq:a6c}
a_6^c(\nu) = n_0\dfrac{1 + n_1\nu + n_2\nu^2 + n_3\nu^3}{1+d_1\nu} \ ,
\ee
where the coefficients $(n_0,n_1,n_2,n_3,d_1)$ are given by Eqs.(4.3)-(4.7) of~\cite{Nagar:2019wds}.
The (point-wise) determination of this function relies on EOB/NR time-domain phasing comparisons. 
For each selected value of $q$, $a_6^c$ is varied manually until the EOB/NR phase agreement is 
smaller than (or of the same order as)  the NR phase uncertainty at merger. For SXS this (probably conservative) 
error is estimated by taking the difference  between two resolutions at merger time. 
This is done using data extrapolated at infinity with $N=3$ order.
For example, for $q=7$, Ref.~\cite{Nagar:2019wds} used the SXS:BBH:0298 dataset and the phase 
uncertainty at merger estimated in this way gives $\delta\phi^{\rm NR}_{\rm mrg}=-0.0775$~rad 
(see Table~I of~\cite{Nagar:2019wds}).
Figure~\ref{fig:q7_sxs} shows our current state-of-the-art for $q=7$ and SXS:BBH:0298. With
$a_6^c(\nu)$ given by Eq.~\eqref{eq:a6c} one has $\Delta\phi^{\rm EOBNR}_{22}\sim -0.244$~rad
at merger point, that is of the same order as, but larger than, the corresponding NR 
uncertainty $\delta\phi^{\rm NR}_{\rm mrg}$ mentioned above. Figure~\ref{fig:q7_sxs} is obtained
with $a_6^c(7/64)\approx -25.562$ from Eq.~\eqref{eq:a6c} and delivers an analytic model that
is NR-faithful for any purpose. However, the \teob{} model is robust and flexible enough to allow 
us to be even {\it less} conservative and actually reach the NR-error level mentioned above.
With $a_6^c=-33$ we get a dephasing at merger $\sim -0.06$ rad, as illustrated in Fig.~\ref{fig:q7_sxs_m33}.
\begin{figure*}[t]
	\center
	\includegraphics[width=0.33\textwidth]{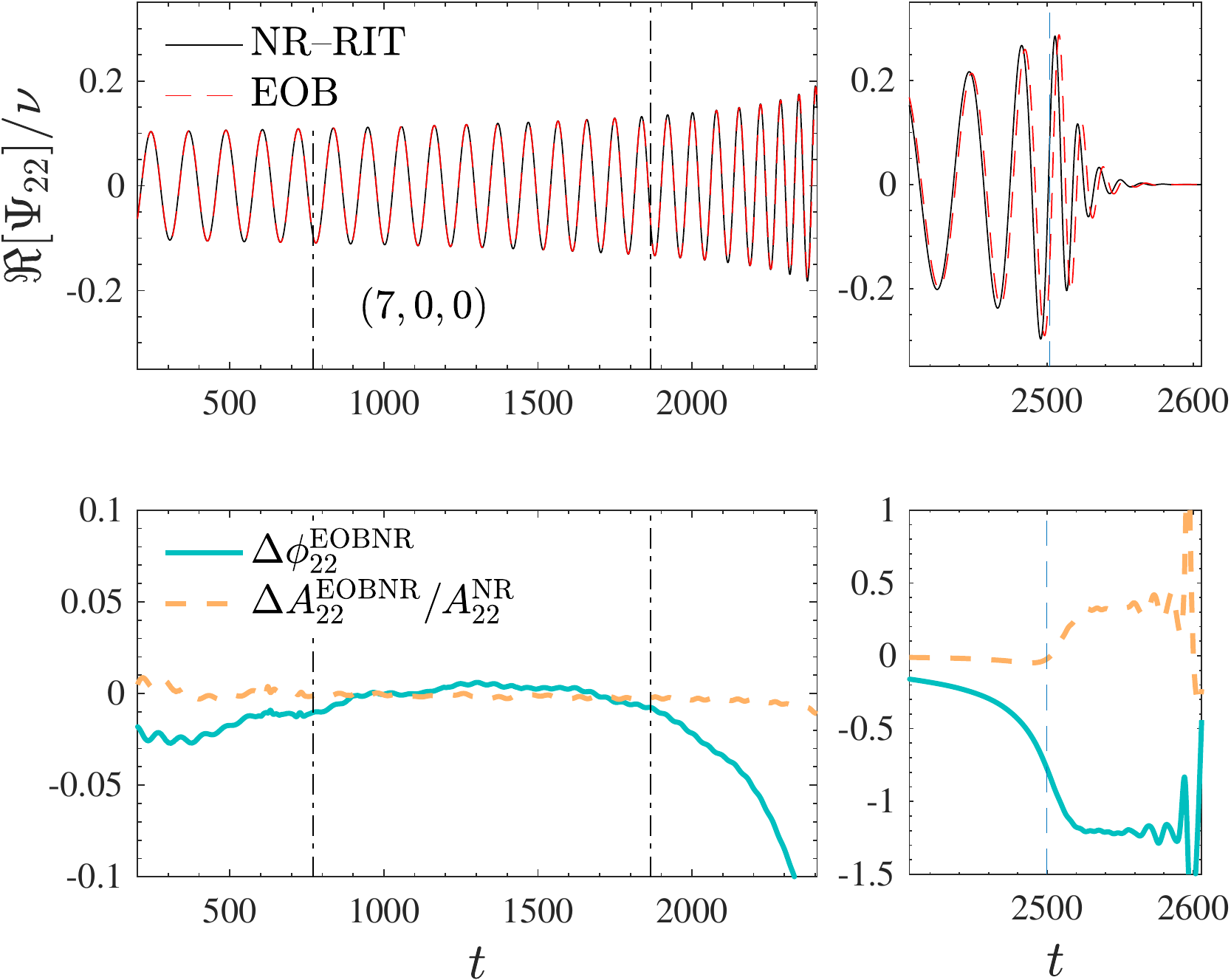}
	\includegraphics[width=0.317\textwidth]{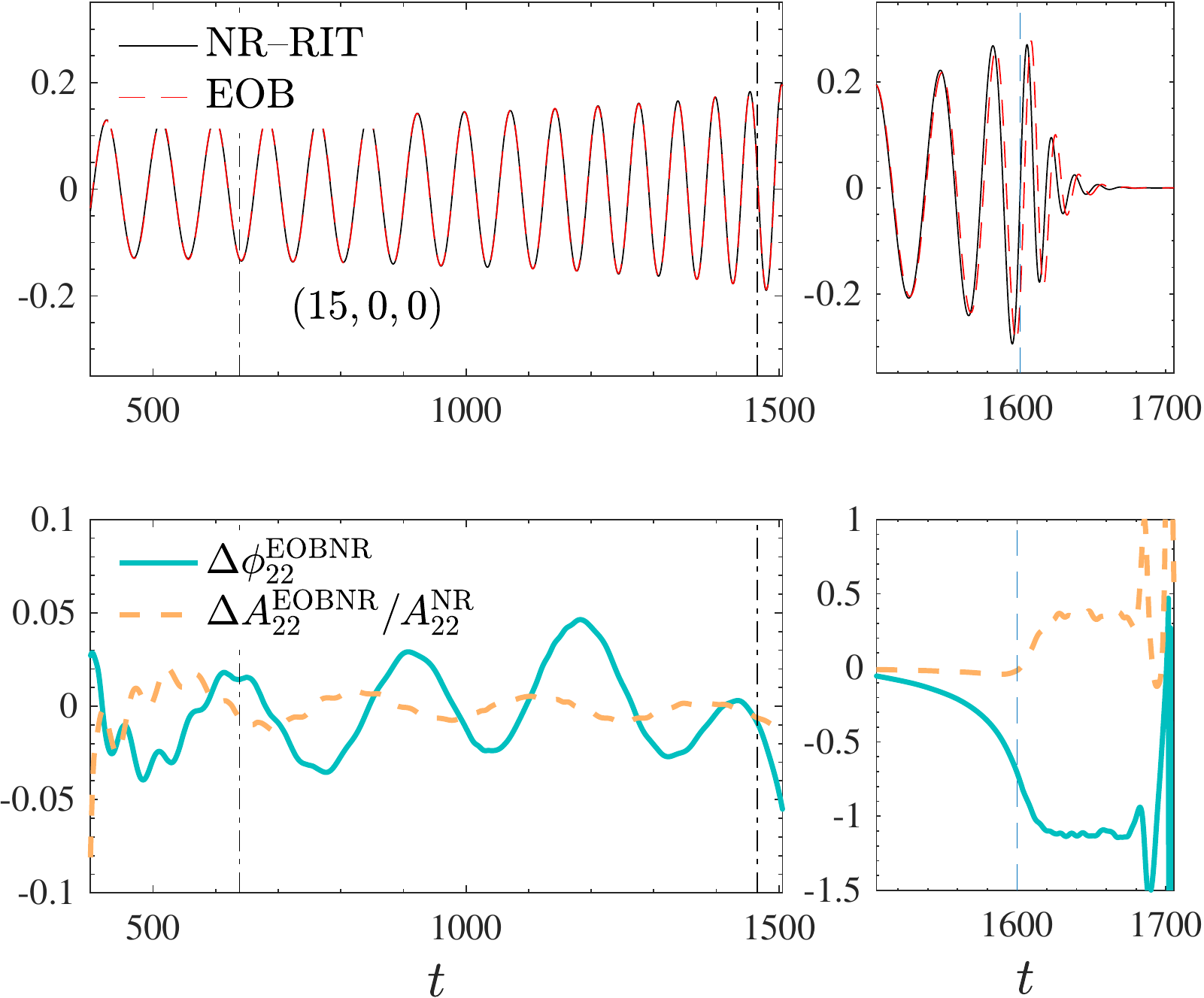}
	\includegraphics[width=0.309\textwidth]{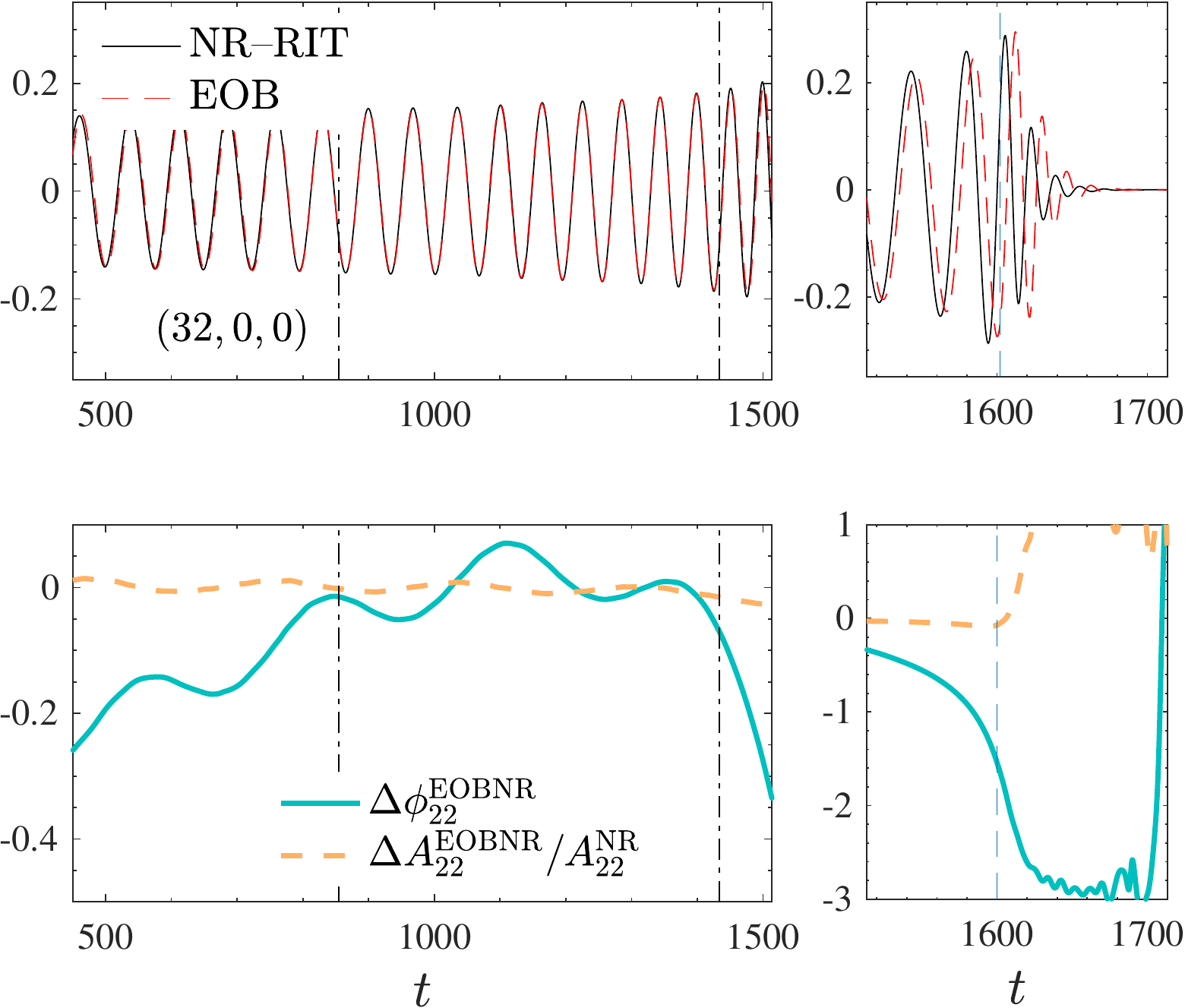}
        \caption{\label{fig:RIT_early} EOB/NR phasing comparison with the waveforms aligned during the early inspiral
        (vertical dash-dotted lines in the plots). We use here resolution-extrapolated waveforms. For $q=15$ and 
        $q=32$ the comparison is clearly affected by residual initial eccentricity in the simulation.}
\end{figure*}
\begin{figure}[t]
	\center
	\includegraphics[width=0.45\textwidth]{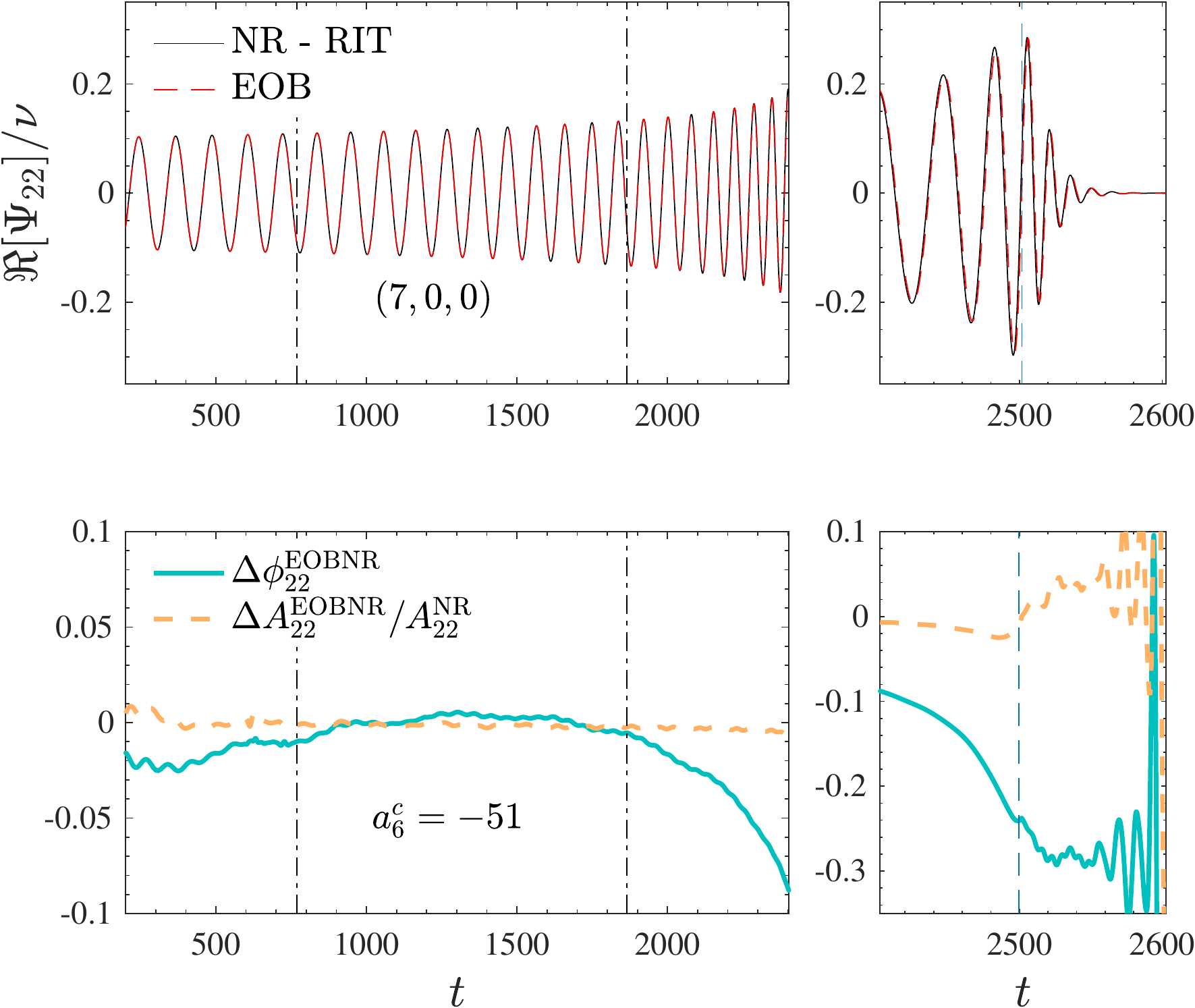}
        \caption{\label{fig:RIT_early_a6c} Attempt of NR-informing $a_6^c$ using the $q=7$ RIT data.
        For $a_6^c=-51$ we can reduce the phase difference around merger at $\sim -0.2$~rad, compatible
        with the NR uncertainty, but this does not affect the phase difference during the inspiral.}
\end{figure}
Comparing Fig.~\ref{fig:q7_sxs} and~\ref{fig:q7_sxs_m33} one sees that the EOB phasing during the long inspiral is
very accurate\footnote{The phase difference oscillates around zero due to the small residual NR eccentricity.} and the change
in $a_6^c$ only affects the last 5 or 6 orbits. On the basis of this analysis,  it seems evident that the level of NR-faithfulness 
that \teob{} can reach {\it depends on} the NR uncertainty. Note in this respect that the SXS simulations were not performed 
with the scope of accurately informing an EOB model. As a consequence, their uncertainties (see e.g. Table~I of Ref.~\cite{Nagar:2019wds}),
obtained by taking the difference of two resolutions, might be either too large or too small for our purposes.
In fact, seen all the complications of NR simulations, the best setup to NR-informing the EOB model would be to have
at hand different, error-controlled, NR-simulations with equivalent length obtained with different numerical methods.
The open question is then to determine to which extent our $a_6^c(\nu)$ function is independent of the choice of NR data.
To attempt an answer, the left panel of Fig.~\ref{fig:RIT_early} shows the EOB/NR phasing comparison for $q=7$ using RIT
NR data extrapolated to infinite resolution. The alignment interval is the same of Fig.~\ref{fig:q7_sxs}, but the phase difference
accumulated up to merger is $\simeq -1$~rad. Given our error estimate on the $q=7$ RIT waveform in Fig.~\ref{fig:rit_resolution},
we expect the NR phase uncertainty to be $\sim 0.3$ rad up to merger. The same conclusion comes from Fig.~\ref{fig:sxs_vs_rit},
where both RIT and SXS waveforms show a high degree of consistency among themselves and with the \teob{} one when aligned
during the late plunge phase. We conclude that the effect in Fig.~\ref{fig:RIT_early} is related to the numerical errors accumulated
during the inspiral, that are probably larger than the SXS ones as already suggested in Fig.~\ref{fig:rit_resolution} above. 
However, on the basis of the complexity of NR simulations and the very different numerical methods employed to obtain the SXS 
and RIT data, it is not a priori completely excluded the existence of subtle systematics on both sides. A similar behavior of the 
phase difference is found also for the $q=32$ and $q=64$ datasets, although it is more compatible with the phase uncertainty.

We conclude this discussion on the dependence of the EOB-tuning on NR data with the following exercise.
Let us suppose to fully trust the $q=7$ RIT data and use them as target to inform $a_6^c$. To do so, we will need a 
new value of $a_6^c$, more negative than the current $-25.562$, so to attempt to absorb the phase difference around merger.
The result of this exercise is shown in Fig.~\ref{fig:RIT_early_a6c}: for $a_6^c=-51$,  the phase difference at 
merger is reduced to $\sim -0.25$~rad, compatible with the NR phase uncertainty. This tuning however
has only a marginal effect on the phase difference during the inspiral, giving thus another 
indication that one needs improvements on the NR side. This suggests that \teob{} is  flexible, because it can be 
tuned to NR when needed, but at the same time {\it rigid and robust}, in the sense that it can be used as an auxiliary tool 
to spot uncertainties (or systematics) in the NR simulations.
This is especially true during the inspiral, where \teob{} performs best.

From our brief analysis it is clear that \teob{} can be robustly informed only using NR simulations with 
very well controlled (and small, $\lesssim 0.1$~rad) numerical uncertainties. If this seems to be true for 
SXS, it is not yet the case for our RIT data, although we clearly proved the consistency between 
the two NR datasets.
However, after this first exploration of the large mass ratio regime and comparisons
of NR waveforms with those of EOB, we have learned that NR can inform
analytic models to improve the fits in this computationally challenging regime. 
The NR waveforms we used were still first explorations, particularly for $q=(32,64,128)$ and
we can now revisit this scenarios with improved accuracy. This is clearly
true also for the $q=7$ case, that would serve as an additional benchmark for
the corresponding SXS dataset and the related NR-informed quantities within \teob{}.
We also note that, as indicated by Figs.~\ref{fig:q7_sxs}
and~\ref{fig:q7_sxs_m33}, shorter NR simulations, with only 10 orbits or less,
might be sufficient for additionally informing \teob{}, provided that they can be
pushed to an accuracy comparable to that of SXS data.
Some of the areas of immediate improvements are: (i) the use of improved 
gauge conditions, as described in Ref.~\cite{Rosato:2021jsq}; (ii) the use 
of different grid structures than in~\cite{Lousto:2020tnb} to emphasize either 
inspiral or ringdown accuracy; (iii) lengthy simulations with global increase 
in the resolution, and (iv) reduction of the initial eccentricity with the methods
of~\cite{Buonanno:2010yk} or~\cite{Walther:2009ng}.

\section{Conclusions}
\label{sec:end}
It is generally believed that a state-of-the-art EOB-based waveform model, specifically \teob{}, is not
robust and trustable outside the so-called {\it domain of calibration}, i.e. that region of the parameter
space covered by the NR simulations that are used to inform the model. We explicitly proved that
this is not true, at least for \teob{}. Specifically, we have used recently published NR 
simulations~\cite{Lousto:2020tnb,Healy:2020iuc,Rosato:2021jsq}  of coalescing BBHs with mass 
ratio from 15 to 128 to validate \teob{} in the large mass-ratio regime. This is the first comparison 
of a semi-analytical waveform model to NR simulations in this corner of the parameter space. 
The excellent mutual consistency we have found between NR and EOB waveform data 
gives additional evidence that \teob{} is currently the most robust, versatile and NR-consistent 
EOB-based, spin-aligned, waveform model available. Our work complements then the
findings of Refs.~\cite{Nagar:2020pcj,Riemenschneider:2021ppj,Albertini:2021tbt}. 

In summary:
\begin{itemize}
\item[(i)] Focusing first on the $(2,2)$ mode, we find an excellent degree of EOB/NR 
consistency during merger and ringdown up to $q=128$;
\item[(ii)] For the inspiral, the numerical 
truncation error increases progressively with the mass ratio. Still, the EOB/NR dephasings we 
find are coherent with the expected NR uncertainty;
\item[(iii)] Similar consistency through merger 
and ringdown is found for all available EOB higher modes, 
$(2,1)$, $(3,3)$, $(3,3)$, $(3,2)$, $(4,4)$ and $(4,3)$,
except for the $(5,5)$ mode.
\item[(iv)] The native implementation of the $(5,5)$  multipole develops unphysical 
features  at merger and during ringdown, which are related to inaccuracies in the NR-informed fits of 
 Ref.~\cite{Nagar:2020pcj}. These features  show up for $q\gtrsim 15$. We thus use the 
 new NR data discussed here to inform an improved $\ell=m=5$ ringdown description. 
 With this new input, the model is tested to be accurate up to $q=128$, and it is smoothly connected with results 
 in the test-particle limit.
The new fit discussed here is implemented in the last public version of \teob{}.
\end{itemize}
Our findings highlight the importance of producing highly accurate NR
simulations that cover the transition to merger and ringdown in all crucial corners of the 
parameter space. It also shows the robustness of the analytical scheme that is used to construct 
the merger-ringdown part of the EOB multipolar waveform~\cite{Damour:2014yha}: once new
NR data are available, one can just use them to improve the NR-informed fits, easily removing
pathological behaviors that may occur in the analytical waveform around merger.
Sparse, but very accurate, NR simulations remain the only tool available to incorporate an accurate
merger-ringdown description within waveform models. We hope that the control of quantities
like $\Delta t_{\lm}$, the delay between the peak of each multipole and the $\ell=m=2$ one,
becomes of primary importance for forthcoming NR simulations.

Let us finally stress that our NR simulations effectively allow us to quantitatively probe {\it only} the plunge, 
merger and ringdown regime of \teob{}. In principle we would need {\it long} simulations with mass ratio 
$q>10$, with a typical SXS accuracy, to probe the radiation-reaction  driven long inspiral. One should however
be aware that the radiation reaction of \teob{} incorporates a large amount of PN information, in resummed
form, in particular {\it hybridizing} $\nu$-dependent terms with test-mass results up to (relative) 
6PN accuracy~\cite{Nagar:2020pcj} for all flux modes up to $\ell=6$. The $\ell=7$ and $\ell=8$ modes, however,
rely on less PN information, and an improvement with test-mass data (following Ref.~\cite{Albanesi:2021rby}) 
could be useful. In general, these improvement to the dissipative sector of the model are expected
to be important for constructing long-inspiral waveform templates for 3G detectors.
By contrast, {\it shorter} NR simulations, $\sim 10$~orbits, with reduced eccentricity and 
accuracy comparable to the SXS ones, would be useful to probe more accurately the full 
transition from late-inspiral  to plunge and merger, possibly informing the EOB dynamics for large mass ratios.
This data would also independently benchmark the NR-informed  EOB interaction potential, that 
currently only relies on strong-field information extracted from SXS simulations. This kind of simulations 
are within reach of our numerical techniques and will be pursued in the future.

\begin{acknowledgments}
  We are grateful to P.~Rettegno, R.~Gamba and M.~Agathos for the implementation
  and testing of the new $\ell=m=5$ fits within both the stand-alone and the LAL
  implementation of \teob{}.
  S.~B. acknowledge support by the EU H2020 under ERC Starting
  Grant, no.~BinGraSp-714626.  
COL and JH gratefully acknowledge the National Science Foundation
(NSF) for financial support from Grants No.\ PHY-1912632.
This work used the Extreme
Science and Engineering Discovery Environment (XSEDE) [allocation
  TG-PHY060027N], which is supported by NSF grant No. ACI-1548562,
and project PHY20007 Frontera,
an NSF-funded petascale computing system
at the Texas Advanced Computing Center (TACC).
Computational resources were also provided by the NewHorizons, BlueSky
Clusters, and Green Prairies at the Rochester Institute of Technology,
which were supported by NSF grants No.\ PHY-0722703, No.\ DMS-0820923,
No.\ AST-1028087, No.\ PHY-1229173, No.\ PHY-1726215, and No.\ PHY-201842.
A.A. has been supported by the fellowship Lumina Quaeruntur No.
LQ100032102 of the Czech Academy of Sciences.
\end{acknowledgments}

\bibliography{refs,local}

\end{document}